\documentclass[aps,pop,twocolumn,showpacs,superscriptaddress,article]{revtex4-1}
\usepackage{mathrsfs}
\usepackage{amsmath}
\usepackage{amsthm}
\usepackage{bm}
\usepackage{lineno}
\usepackage{color}
\usepackage{graphicx}
\usepackage{hyperref}
\usepackage{subfig}
\begin{document}
%\linenumbers

\title{Gyrokinetic Theory of Low-frequency Electromagnetic Waves in Finite-$\beta$ Anisotropic Plasmas}

%\correspondingauthor{Liu Chen}
%\email{liuchen@zju.edu.cn}
%\title{Gyrokinetic Theory of Low-frequency Electromagnetic Waves in Finite-$\beta$ Anisotropic Plasmas}

\author{Haotian Chen}
\affiliation{Institute of Space Science and Technology, Nanchang University, Nanchang, 330031, China}
\affiliation{Department of Atomic, Molecular and Nuclear Physics, University of Seville, Seville, 41012, Spain}
\author{Liu Chen}\email[]{Email: liuchen@zju.edu.cn}
\affiliation{Institute for Fusion Theory and Simulation and Department of Physics, Zhejiang University, Hangzhou, 310027, China}
\affiliation{Department of Physics and Astronomy, University of California, Irvine, California 92697, USA}

\begin{abstract}
	We present a gyrokinetic theory for the electromagnetic waves and instabilities  with frequencies much lower than the ion cyclotron frequency in finite-$\beta$ anisotropic uniform plasmas. Here, $\beta$ is the ratio between plasma and magnetic pressures.
	Kinetic effects due to both the finite Larmor radii and   wave-particle resonances are fully kept in the analysis.
	Corresponding linear dispersion relation and wave polarizations, valid for general $\beta$ value and perpendicular wavelength, are then specifically derived for a bi-Maxwellian plasma.
	Analytic expressions for the criteria of kinetic firehose and mirror instabilities are also given.
	The mode frequency, stability, and wave polarization of a broad spectrum of normal modes  are then investigated numerically in a systematic study over a set of parameters. 
	Our study clearly demonstrates that, due to the finite ion Larmor radius effect, the ion-sound wave, mirror mode  and  shear Alfv\'{e}n wave are intrinsically coupled.
\end{abstract}

\pacs{52.30.Gz, 52.35.Bj, 52.35.Py, 96.50.Ci}
\maketitle

%\begin{linenumbers}
\section{Introduction}
\label{sec:may:01:16:01}
Electromagnetic fluctuations with frequency much lower than the ion cyclotron frequency, such as magnetohydrodynamic (MHD) Alfv\'{e}n waves,  are fundamental to magnetized plasma dynamics. 
%Among those waves are the ion-sound wave (ISW), kinetic Alfv\'{e}n wave (KAW)  and mirror mode (MM).
They are prevalent in space, astrophysical and laboratory plasmas and play important roles in heating, acceleration and transport of charged particles.
For example, according to the double-adiabatic CGL equation \citep{chew}, the geometric expansion of solar wind in the spiral structured interplanetary magnetic field will result in large temperature anisotropies near 1 AU \citep{phillips,kasper}.
The \emph{in situ} observations, however, have shown that the solar wind  at this distance is close to the isotropic state \citep{marsch82, phillips}. This raises the question of solar wind isotropy.
Naturally, the pressure-anisotropy-driven instabilities, such as firehose and mirror instabilities, are expected to be important contributors to isotropization processes, and have been investigated extensively in both theoretical studies (e.g.,\citep{yoon,hellinger07,kunz,yoon19}) and satellite observations (e.g.,\citep{kasper, hellinger06,stverak, bale,chen15}).
Theoretically, most treatments available in the literature are based on either the  fluid  or drift-kinetic descriptions.
The fully kinetic analysis, on the other hand, often involves the complicated procedures of taking the low-frequency limit of Vlasov theory \citep{yoon}, which makes the model analytically difficult to treat.

In this work, we adopt the linear gyrokinetic theory \citep{antosen,catto81,frieman} to explore the wave and stability properties of low-frequency electromagnetic fluctuations, i.e.,
the ion-sound wave (ISW), the shear Alfv\'{e}n wave (SAW)/kinetic Alfv\'{e}n wave (KAW) excited by firehose instability, and the mirror mode (MM), in finite-$\beta$ anisotropic uniform plasmas. Here, $\beta$ is the ratio between plasma and magnetic pressures. Unlike the most of previous studies, the present theoretical framework includes full finite Larmor radius (FLR) effect and wave-particle resonance, while the resultant model is still analytical tractable and offers a useful tool for gaining insights into the underlying physics.

This paper is organized as follows. Section \ref{sec:oct:15:01:54} presents the gyrokinetic theoretical model. In Sec. \ref{sec:oct:15:08:18}, we adopt the bi-Maxwellian distributions, derive the general linear wave equations and analytically investigate the corresponding wave and stability properties.
Section \ref{sec:nov:07:19:46} contains the numerical results of a systematic parameter study. Final conclusions and discussions are given in Sec. \ref{sec:oct:01:16:05}.
%A derivation of the instability threshold, which is valid for kinetic waves under certain circumstances,  is present in Appendix \ref{sec:nov:30:10:39}.
%\end{linenumbers}
%\begin{linenumbers}

\section{Gyrokinetic Theory}
\label{sec:oct:15:01:54}
For simplicity and, hence, clarity, we consider a uniform and finite-$\beta$ plasma slab immersed in a straight shearless uniform background magnetic field $\bm{B}=B\bm{e}_{z}$. 
Following \citep{chen91}, the low-frequency plasma fluctuations could be described by three fluctuating scalar fields: the electrostatic potential $\delta\phi$, the parallel component of the vector potential $\delta A_{\parallel}$ and the compressional component of the magnetic field fluctuation $\delta B_{\parallel}$. For convenience, we may further introduce the scalar induced potential $\delta\psi$ as
\begin{eqnarray}
\label{eq:psi}
	-\nabla_{\parallel}\delta\psi=\frac{1}{c}\partial_{t}\delta A_{\parallel}.
\end{eqnarray}
The parallel electric field is, thus, given by $\delta E_{\parallel}=-\nabla_{\parallel}(\delta\phi-\delta\psi)$.
%The ideal MHD limit, i.e., the parallel electric field fluctuation $\delta E_{\parallel}=0$, is thereby recovered for $\delta\psi=\delta\phi$.
With this representation, we can adopt the fluctuating fields $(\delta\phi,\delta\psi,\delta B_{\parallel})$ as our variables and write them as superposition of plane waves, namely,
\begin{eqnarray}
\label{eq:fourier}
	[\delta\phi,\delta\psi,\delta B_{\parallel}]=\sum_{\bf{k}}[\delta\phi_{k},\delta\psi_{k},\delta B_{\parallel,k}]e^{i(\bf{k}\cdot\bf{x}-\omega t)}.
\end{eqnarray}
In terms of the guiding-center variables $(\varepsilon=v^{2}/2,\mu=v_{\perp}^{2}/2B,\alpha,\hat{\sigma}=\textrm{sign}(v_{\parallel}))$ with $\alpha$ the gyrophase, the perturbed distribution function $\delta F_{k}$, by separating the adiabatic and convective particle responses \citep{chen91}, can be expressed as
\begin{eqnarray}
\label{eq:deltaf}
	\delta F_{ks}&=&\delta K_{ks} e^{iL_{k}}+\frac{q_{s}}{m_{s}}\{\delta\phi_{k}\frac{\partial F_{0s}}{\partial\varepsilon}-\frac{Q}{\omega}F_{0s}J_{0}\delta\psi_{k} e^{iL_{k}}\nonumber\\
	& &+\frac{\partial F_{0s}}{B\partial \mu}[(\delta\phi_{k}-\frac{k_{\parallel}v_{\parallel}}{\omega}\delta\psi_{k})(1-J_{0}e^{iL_{k}})\nonumber\\
	& &-\frac{v_{\perp} J_{1}}{k_{\perp}c}\delta B_{\parallel k}e^{iL_{k}}]\}.
\end{eqnarray}
Here, the subscript $s$ stands for particle species, $Q F_{0s}=\omega\partial_{\varepsilon}F_{0s}$,
$J_{n}(\lambda)$ is a Bessel function of $n$-index, $\lambda=k_{\perp}v_{\perp}/\omega_{c}$ and $L_{k}=\bm{k}\cdot\bm{b}\times\bm{v}/\omega_{c}$ with $\omega_{c}$ being the cyclotron frequency.
The kinetic compression response $\delta K_{ks}$, meanwhile, satisfies the gyrokinetic equation \citep{catto81,chen91}
\begin{eqnarray}
\label{eq:gke}
	(\hat{\sigma}|v_{\parallel}|k_{\parallel}-\omega)\delta K_{ks}=i\frac{q_{s}}{m_{s}}QF_{0s}\delta S_{k},\nonumber\\
\end{eqnarray}
where
\begin{eqnarray}
\label{eq:source}
	\delta S_{k}=J_{0}(\delta\phi_{k}-\delta\psi_{k})+\frac{v_{\perp} J_{1}}{k_{\perp}c}\delta B_{\parallel k},
\end{eqnarray}
corresponds to the effective parallel force in the present model.

To determine electromagnetic fields with the standard linear gyrokinetic ordering \citep{rutherford,taylor,frieman,howes06}, namely,
\begin{eqnarray}
\label{eq:ordering}
	|\frac{\omega}{\omega_{c}}|\sim |\frac{k_{\parallel}}{k_{\perp}}|\ll 1;\quad |k_{\perp}\rho|\sim\mathcal{O}(1);\quad \beta\sim\mathcal{O}(1),
\end{eqnarray}
with $\rho$ denoting the Larmor radius,
we employ the quasineutrality condition for $\delta\phi_{k}$, 
\begin{eqnarray}
\label{eq:quasineutrality}
	0&=&\sum_{s}q_{s}\langle J_{0}\delta K_{k s}\rangle+\sum_{s}\frac{q_{s}^{2}}{m_{s}}\langle\{\delta\phi_{k}\frac{\partial F_{0s}}{\partial\varepsilon}-\frac{Q}{\omega}F_{0s}J^{2}_{0}\delta\psi_{k} \nonumber\\
	& &+\frac{\partial F_{0s}}{B\partial \mu}[(1-J^{2}_{0})(\delta\phi_{k}-\frac{k_{\parallel}v_{\parallel}}{\omega}\delta\psi_{k})\nonumber\\
	& &-\frac{v_{\perp} J_{0}J_{1}}{k_{\perp}c}\delta B_{\parallel k}]\}\rangle,
\end{eqnarray}
where $\langle \cdots\rangle$ denotes the integration in velocity space.
The equation for the compressional component of magnetic field perturbation $\delta B_{\parallel k}$ is obtained from the perpendicular component of Ampere's law,
\begin{eqnarray}
\label{eq:ampere}
	\delta B_{\parallel k}&=&\frac{4\pi}{ck_{\perp}}\sum_{s}q_{s}\langle v_{\perp} J_{1}\{\frac{q_{s}}{m_{s}}[\frac{Q}{\omega}F_{0s}J_{0}\delta\psi+\frac{\partial F_{0s}}{B\partial\mu}(J_{0}\nonumber\\
	& &\times(\delta\phi-\frac{k_{\parallel}v_{\parallel}}{\omega}\delta\psi_{k})+\frac{v_{\perp}J_{1}\delta B_{\parallel k}}{k_{\perp}c})]-\delta K_{k s}\}\rangle.
\end{eqnarray}
It can be readily noted that Eq. (\ref{eq:ampere}) is formally equivalent to the perpendicular pressure balance condition $\nabla_{\perp}(B\delta B_{\parallel}+4\pi \delta P_{\perp})=0$ to the lowest order in $\mathcal{O}(|k_{\parallel}|/|k_{\perp}|)$.
Therefore, the fast magnetoacoustic waves have been suppressed within the low-frequency gyrokinetic theory by balancing the perpendicular plasma pressure with the magnetic field pressure \citep{chen16}.
The equation for the perturbed field $\delta\psi_{k}$, on the other hand, is written in terms of the following vorticity equation \citep{chen16}, which is derived from the quasineutrality condition, $\nabla\cdot\delta\bm{J}=0$, along with the parallel Ampere's law,
\begin{eqnarray}
\label{eq:vorticity}
	0&=&\frac{ c^{2}k_{\parallel}^{2}}{4\pi\omega^{2}}k_{\perp}^{2}\sigma_{k}\delta\psi_{k}+\sum_{s}\frac{q_{s}^{2}}{m_{s}}\{\langle \frac{Q}{\omega} F_{0s}(1-J^{2}_{0})\rangle\delta\phi_{k}\nonumber\\
	& &+\langle\frac{\partial F_{0s}}{B\partial \mu}[(\delta\phi_{k}-\frac{k_{\parallel}v_{\parallel}}{\omega}\delta\psi_{k})(1-J^{2}_{0})-\frac{v_{\perp} J_{0}J_{1}}{k_{\perp}c}\delta B_{\parallel k}]\rangle\nonumber\\
	& &-\langle  \frac{Q}{\omega} F_{0s} \frac{v_{\perp}J_{0} J_{1}}{k_{\perp} c}\rangle\delta B_{\parallel k},
\end{eqnarray}
where
\begin{eqnarray}
\label{eq:sigma}
	\sigma_{k}=1+\frac{4\pi}{k_{\perp}^{2}c^{2}}\sum_{s}\frac{q_{s}^{2}}{m_{s}}\langle v_{\parallel}^{2}(1-J_{0}^{2})\frac{\partial F_{0s}}{B\partial\mu}\rangle,
\end{eqnarray}
is the kinetic firehose stability term including finite Larmor radius  corrections in an anisotropic distribution.
Equations (\ref{eq:quasineutrality}), (\ref{eq:ampere}) and (\ref{eq:vorticity}), along with the linear gyrokinetic equation (\ref{eq:gke}),  form a closed set of rather general equations governing the linear low-frequency wave dynamics in finite-$\beta$ anisotropic uniform plasmas.

\section{General Wave Properties in a Bi-Maxwellian Plasma}
\label{sec:oct:15:08:18}
To make further analytical progress and gain useful insights into the linear wave and stability properties, we assume that the anisotropic equilibrium distribution functions are bi-Maxwellian:
\begin{eqnarray}
\label{eq:F0}
	F_{0s}=\frac{N_{s}}{\pi^{3/2} v^{2}_{ts\perp} v_{ts\parallel}}e^{-\frac{v_{\parallel}^{2}}{v_{ts\parallel}^{2}}-\frac{v_{\perp}^{2}}{v_{ts\perp}^{2}}},
\end{eqnarray}
where $N_{s}$ is the equilibrium density, $v_{ts\perp(\parallel)}$ denotes the perpendicular (parallel) thermal velocity and $T_{s\perp(\parallel)}=m_{s}v_{ts\perp(\parallel)}^{2}/2$ are the corresponding temperatures.
Substituting Eq. (\ref{eq:F0}) into the Eqs. (\ref{eq:gke}-\ref{eq:vorticity}), one can, after some straightforward  algebra, derive the following normal mode equations,
\begin{eqnarray}
\label{eq:eigenmode}
	Q_{1}\Phi_{\parallel}+V_{1}\Psi+Q_{3}B_{\parallel}&=&0\nonumber\\
	V_{1}\Phi_{\parallel}+(V_{1}+\frac{V_{2}}{\Omega^{2}})\Psi+V_{3}B_{\parallel}&=&0\\
	-\frac{\beta_{i\parallel}}{2}(Q_{3}\Phi_{\parallel}+V_{3}\Psi)+A_{3}B_{\parallel}&=&0.\nonumber
\end{eqnarray}
Here, for convenience, we have dropped the subscript $k$, and adopted the following normalizations
\begin{eqnarray}
\label{eq:normalization}
	\Phi=\frac{2e\delta\phi}{m_{i}v_{ti\parallel}^{2}},\quad \Psi=\frac{2e\delta\psi}{m_{i}v_{ti\parallel}^{2}},\quad \frac{\delta B_{\parallel}}{B}=B_{\parallel}
\end{eqnarray}
and $\Phi_{\parallel}=\Phi-\Psi$, as noted above, is related to the parallel electric field. 
Meanwhile, the coefficients are given as
\begin{eqnarray}
\label{eq:coef}
	Q_{1}&=&-\sum_{s}\frac{T_{i\parallel}}{T_{s\parallel}}[ (1+\xi_{s}Z_{s} \Gamma_{0s})+a_{s}(1-\Gamma_{0s})],\nonumber\\
	Q_{3}&=&\sum_{s}\frac{|q_{s}|}{q_{s}}\frac{\Gamma_{0s}-\Gamma_{1s}}{1+a_{s}}(a_{s}-\xi_{s}Z_{s}),\nonumber\\
	V_{1}&=&-\sum_{s}(1+a_{s})\frac{T_{i\parallel}}{T_{s\parallel}}(1-\Gamma_{0s}),\nonumber\\
	V_{2}&=&\sigma_{k}(1+a_{i})b_{i},\nonumber\\
	V_{3}&=&\sum_{s}\frac{|q_{s}|}{q_{s}}(\Gamma_{0s}-\Gamma_{1s}),\nonumber\\
	A_{3}&=&-1+\sum_{s}\frac{\beta_{s\perp}}{1+a_{s}}(\Gamma_{0s}-\Gamma_{1s}) [\xi_{s}Z_{s}-a_{s}],
\end{eqnarray}
where $T_{e\parallel}/T_{i\parallel}=\tau$ is the parallel temperature ratio between electron and ion, $\beta_{s\perp(\parallel)}=8\pi N_{0}T_{s\perp(\parallel)}/B^{2}$, $\xi_{s}=\omega/|k_{\parallel}|v_{ts\parallel}$, $\Omega=\omega/|k_{\parallel}|v_{A}$, $q_{s}$ is the particle charge, $v_{A}=(B^{2}/4\pi N_{0}m_{i})^{1/2}$ is the Alfv\'{e}n speed and $a_{s}=(T_{s\parallel}^{2}/T_{s\perp}^{2})-1$ explicitly accounts for  the temperature anisotropy.
$Z_{s}=Z(\xi_{s})$ denotes the usual plasma dispersion function.
	$\Gamma_{j s}=I_{j}(b_{s}) \textrm{exp}(-b_{s})$ can be regarded as FLR effect with $I_{j}$ being the first kind modified Bessel function and $b_{s}=k_{\perp}^{2}\rho_{ts\perp}^{2}/2$.
	The gyrokinetic firehose stability term, Eq. (\ref{eq:sigma}),  becomes $\sigma_{k}=1-\sum_{s}\beta_{s\perp}a_{s}(1-\Gamma_{0s})/2 b_{s}$. 
	Incidentally, it is worthwhile mentioning that $Q_{1}$, $V_{1}+V_{2}/\Omega^{2}$ and $A_{3}$ correspond, respectively, to the ISW, SAW/KAW and MM.
The complex normal mode frequency $\Omega$ can then be obtained from the characteristic equation of Eq. (\ref{eq:eigenmode}), i.e., the following linear dispersion relation
\begin{eqnarray}
\label{eq:char}
	D&\equiv&\Omega^{2}\{V^{2}_{1}A_{3}+\frac{\beta_{i\parallel}}{2}V_{3}[2Q_{3}V_{1}-V_{3}Q_{1}]\}\nonumber\\
	& &-(\Omega^{2}V_{1}+V_{2})[Q_{1}A_{3}+\frac{\beta_{i\parallel}}{2}Q_{3}^{2}]=0.
\end{eqnarray}
%Comparing Eq. (\ref{eq:char}) with the dispersion relation derived by \citep{yoon}, it is seen that the high-order corrections due to finite $k_{\parallel}^{2}v_{A}^{2}/\omega_{ci}^{2}$  have been neglected here, consistent with the gyrokinetic ordering, Eq. (\ref{eq:ordering}).

Before proceeding further into detailed analyses, we first remark on some general spectral properties of Eq. (\ref{eq:char}).
Note that, due to the symmetry property of the plasma dispersion function \citep{fried} 
\begin{eqnarray}
\label{eq:zpro1}
[Z(\xi)]^{*}=-Z(-\xi^{*}),
\end{eqnarray}
then it can be readily shown that, given a normal mode frequency $\Omega$ of Eq. (\ref{eq:char}), its real conjugate $-\Omega^{*}$ satisfies $D^{*}=0$ and, hence, is also a self-consistent normal mode frequency.
This result is of crucial importance in limiting the distribution of normal mode frequencies.
For example, normal modes exist in pairwise for $Re(\Omega)\ne 0$ modes;
for the $Re(\Omega)=0$ case, e.g., the mirror mode, the real conjugate of $\Omega$ is itself. Since the multiplicity of solutions of Eq.(\ref{eq:char}) is independent of parameters, simple $Re(\Omega)=0$ normal modes will stay in the imaginary axis with varying parameters,
i.e., mirror mode can not have real frequency in the present uniform-plasma analyses.
However, the  reverse process is possible when two $Re(\Omega)\ne 0$ simple normal modes merge and form a solution of multiplicity $2$ at the point $\Omega=0$.
As illustrated later in Figs. (\ref{eps:fh_ai_eig}) and (\ref{eps:fh_ae_eig}), this feature is observed as KAWs are destabilized by the firehose instability. 
%physically it can be interpreted as a reactive process.

We now explore further the stability properties of the linear normal modes given by Eq.(\ref{eq:char}). 
From Eq. (\ref{eq:char}), we note that a self-consistent route for instability to arise in the normal mode frequency is via $\Omega=0$.
Furthermore, noting that $D(|\Omega|\to \infty)<0$ and $\textrm{Im}(D)=0$ at $\Omega=0$, a straightforward Nyquist analysis yields that instability sets in when $\textrm{Re}[D(\Omega=0)]>0$.
We then readily derive the following linear instability condition; that is, at $\Omega=0$, 
\begin{eqnarray}
\label{eq:critical}
	\sigma_{k}[Q_{1}A_{3}+\frac{\beta_{i\parallel}}{2}Q_{3}^{2}]\le 0.
\end{eqnarray}
Note that, as no additional assumption has been made, Eq. (\ref{eq:critical}), thus, provides  a rather general instability condition, valid within the gyrokinetic ordering for bi-Maxwellian uniform anisotropic plasmas.
More specifically, noting that the factor $\sigma_{k}$ corresponds to the firehose instability, the onset of unstable KAWs is given by $\sigma_{k}<0$, i.e.,
\begin{eqnarray}
\label{eq:firehose}
	1-\sum_{s}\frac{a_{s}\beta_{s\perp}}{2b_{s}}(1-\Gamma_{0s})<0.
\end{eqnarray}
 Equation (\ref{eq:firehose}) indicates the well-known necessary instability condition, $a_{s}>0$ ($T_{s\parallel }>T_{s\perp }$), and also demonstrates that the FLR effect is stabilizing for the kinetic firehose instability.
Meanwhile, the $Q_{1}A_{3}+\beta_{i\parallel}Q_{3}^{2}/2$ term in Eq. (\ref{eq:critical}) corresponds to the mirror mode coupled to the ion-sound wave, and the kinetic mirror instability sets in when
\begin{eqnarray}
\label{eq:ms}
	1+\sum_{s}\frac{a_{s}\beta_{s\perp}}{1+a_{s}}(\Gamma_{0s}-\Gamma_{1s})<\frac{-\frac{\beta_{i\parallel}}{2}[\sum_{s}\frac{|q_{s}|}{q_{s}}\frac{a_{s}(\Gamma_{0s}-\Gamma_{1s})}{1+a_{s}}]^{2}}{\sum_{s}\frac{T_{i\parallel}}{T_{s\parallel}} [1+a_{s}(1-\Gamma_{0s})]}.
\end{eqnarray}
Again, Eq. (\ref{eq:ms}) indicates the well-known $a_{s}<0$ ($T_{s\parallel}<T_{s\perp}$) necessary condition for the mirror mode instability.
Meanwhile, the right-hand-side term is due to the coupling to  the ion-sound wave, resulting in an enhancement of the critical $\beta$ threshold.

To gain more analytical insights, we note that the general $3\times 3$ matrix can be further reduced into a $2\times 2$ system in two limits; the low-$\beta$ limit and the long wavelength limit. 
%For simplicity, here we limit our discussion to waves in an isotropic plasma with $a_{s}=0$.

\subsection{Low-$\beta$ limit}
In the low-$\beta$ regime ($m_{e}/m_{i}\ll \beta\ll 1$), the perpendicular Ampere's law Eq. (\ref{eq:ampere}) indicates that the compressional component $B_{\parallel}$ can be self-consistently neglected, and the system is stable, according to Eqs. (\ref{eq:firehose}) and (\ref{eq:ms}). The dispersion relation (\ref{eq:char}) is then reduced to
\begin{eqnarray}
\label{eq:lowbetadisper}
	& &[ (1+\xi_{i}Z_{i} )\Gamma_{0i}+ (1+\xi_{e}Z_{e} ) \frac{\Gamma_{0e}}{\tau}](\Omega^{2}-\frac{b_{i}}{R_{0}})\nonumber\\
	&=&(1+a_{i})b_{i},
\end{eqnarray}
where
\begin{eqnarray}
\label{eq:R0}
	R_{0}=(1-\Gamma_{0i})+\frac{1+a_{e}}{\tau(1+a_{i})}(1-\Gamma_{0e})
\end{eqnarray}
is positive definite. We can easily recognize that Eq. (\ref{eq:lowbetadisper}) describes the coupling between the shear Alfv\'{e}n wave and ion-sound wave due to the ion FLR effect.

More specifically, in the low-frequency region with $|\xi_{e}|^{2}\ll|\Omega|^{2} \ll 1$, Eq. (\ref{eq:lowbetadisper}) can be cast into
\begin{eqnarray}
\label{eq:lowbetadisperlowfre}
	 \xi_{i}Z_{i} =-1-\frac{(1+a_{i})R_{0}}{\Gamma_{0i}}-\frac{\Gamma_{0e}}{\tau\Gamma_{0i}},
\end{eqnarray}
which describes the ion-sound wave branch with $\Psi\simeq 0$.
Noting that the value of the right-hand side of Eq. (\ref{eq:lowbetadisperlowfre}) is smaller than $-1$, thus $|\xi_{i}|^{2}\gtrsim 1$ is expected for the ISWs in the low-$\beta$  regime.
%\begin{eqnarray}
%\label{eq:lowbetadispersound}
	 %\frac{\xi_{i}^{2}}{\tau} =\frac{\Gamma_{0i}}{2(\tau R_{0}+\Gamma_{0e})}.
%\end{eqnarray}

When considering shear/kinetic Alfv\'{e}n waves with $|\xi_{e}|^{2}\ll|\Omega|^{2}\sim1 \ll |\xi_{i}|^{2}$, Eq. (\ref{eq:lowbetadisper}) readily recovers the well-known KAW dispersion relation \citep{hasegawa75, hasegawa76,chen20}
\begin{eqnarray}
\label{eq:lowbetadisper1}
	\Omega^{2}=\frac{b_{i}}{R_{0}}+\frac{\tau (1+a_{i}) b_{i}}{\Gamma_{0e}(1-2\xi_{e}^{2}+i\sqrt{\pi}\xi_{e}e^{-\xi_{e}^{2}})},
\end{eqnarray}
and the corresponding wave polarization
\begin{eqnarray}
\label{eq:lowbetadisper2}
	\frac{\Psi}{\Phi_{\parallel}}=-1-\frac{(1-2\xi_{e}^{2}+i\sqrt{\pi}\xi_{e}e^{-\xi_{e}^{2}})\Gamma_{0e}}{\tau(1+a_{i})R_{0}}.
\end{eqnarray}
From Eq. (\ref{eq:lowbetadisper2}), it is clear that the fields $\Phi_{\parallel}$ and $\Psi$ of KAW are out of phase in the adiabatic electron  limit with $|\xi_{e}|\to 0^{+}$, $\Psi$ dominates in the long wavelength regime, and a finite parallel electric field is induced by coupling with ISW, due to the finite ion Larmor radius (FILR).

%In the high-frequency region with $1\ll|\Omega|^{2}\ll |\xi_{i}|^{2}$, on the other hand, the ion Landau damping effect is negligible and fluctuations are characterized by
%\begin{eqnarray}
%\label{eq:lowbetadisperhighfre}
	%1+\xi_{e}Z_{e}\simeq 0,
%\end{eqnarray}
%and $\Phi_{\parallel}\simeq -\Psi$, i.e., $\Phi\simeq 0$, which can be identified as a gyrokinetic version of plasma oscillation \citep{lee}.

\subsection{Long wavelength limit}
Considering the long wavelength limit $b_{e}\ll b_{i}\ll 1$, we obtain $V_{1}\simeq V_{3}\simeq 0$, and the incompressible SAW is decoupled from the ISW and compressional magnetic fluctuation dynamics. 
As a consequence, one recovers the fluid limit of the SAW normal mode;
\begin{eqnarray}
\label{eq:firehoselw}
	\Omega^{2}=1-\frac{1}{2}a_{i}\beta_{i\perp}-\frac{1}{2}a_{e}\beta_{e\perp},
\end{eqnarray}
which accounts for the firehose instability in high-$\beta$ and $T_{s\parallel}>T_{s\perp}$ environments.
Once the critical anisotropy has been reached,  the normal mode of multiplicity $2$ at the point $\Omega=0$ will be divided into two simple normal modes, one is purely growing and the other is purely damped.
Again, due to the multiplicity being independent of parameters, the simple unstable solution of KAW cannot become a pair of unstable normal modes without going through the $\Omega=0$ point, even if a finite $b_{i}$ is included. Therefore, the instability condition of SAW/KAW is indeed set by Eq. (\ref{eq:critical}), and the unstable SAW/KAW has no real frequency in the present model.

%Furthermore, notice also that 

Meanwhile, another low-frequency branch also exists, characterized by coupling of the ISW to MM, which could be described by combining the quasineutrality and perpendicular Ampere's law. The resultant linear dispersion relation is
\begin{eqnarray}
\label{eq:mmisw}
	Q_{1}A_{3}+\frac{\beta_{i\parallel}}{2}Q_{3}^{2}=0,
\end{eqnarray}
along with the polarization
\begin{eqnarray}
\label{eq:polar_mmisw}
	\frac{B_{\parallel}}{\Phi_{\parallel}}=\sqrt{-\frac{\beta_{i\parallel}Q_{1}}{2A_{3}}}.
\end{eqnarray}
Equation (\ref{eq:mmisw}) recovers the ISW ($Q_{1}\simeq 0$) and MM ($A_{3}\simeq 0$) as limiting cases of, respectively,  $|\beta_{i\parallel}Q_{3}^{2}/2A_{3}|\ll 1$ and $|\beta_{i\parallel}Q_{3}^{2}/2Q_{1}|\ll 1$. 
It is worthwhile mentioning that, from  the $Q_{1}$ term in Eq. (\ref{eq:coef}), the ISW frequency is more sensitive to $\tau=T_{e\parallel}/T_{i\parallel}$.
%This distinctive feature can be used to identify ISWs.

Further analytical progress can be made for MMs by taking the $A_{3}=0$ limit, i.e.,
\begin{eqnarray}
\label{eq:mm1}
	 \frac{\beta_{i\perp}}{1+a_{i}}\xi_{i}Z_{i}+\frac{\beta_{e\perp}}{1+a_{e}}\xi_{e}Z_{e}=1+\frac{a_{i}\beta_{i\perp}}{1+a_{i}}+\frac{a_{e}\beta_{e\perp}}{1+a_{e}}.
\end{eqnarray}
The fact that $\xi Z(\xi)>-1$ is a monotonically decreasing function along the imaginary axis allows us to readily find that there is exactly one pure imaginary solution to Eq.(\ref{eq:mm1}). 
Therefore, the so-called ion and electron mirror modes \citep{pantellini,pokhotelov} are essentially the same mode driven by different anisotropic plasma components.
As a consequence, mirror mode marginal stability also occurs at $\Omega=0$, and the instability condition is given by Eq. (\ref{eq:critical}).
The classical mirror mode dispersion relation \citep{tajiri,hasegawa69}, meanwhile, can be derived by assuming $|\xi_{e}|^{2}\ll |\xi_{i}|^{2}\ll 1$, i.e.,
\begin{eqnarray}
\label{eq:mm2}
	i\sqrt{\pi}\frac{\beta_{i\perp}}{1+a_{i}}\xi_{i}=1+\frac{a_{i}\beta_{i\perp}}{1+a_{i}}+\frac{a_{e}\beta_{e\perp}}{1+a_{e}},
\end{eqnarray}
%the mirror instability can occur in the presence of a negative anisotropy $a_{s}$.
%In the limiting case with the instability drive $a_{s}\to -1$, Eq. (\ref{eq:mmisw}) readily yields that
In general, it should be emphasised that, as indicated by Eq. (\ref{eq:ms}) or Eq. (\ref{eq:mmisw}) ,  the MMs and ISWs are strongly coupled due to the finite $Q_{3}$.
It is interesting to note that, for the electron mirror mode with $|\xi_{e}|\sim 1$, the dispersion relation,  Eq. (\ref{eq:mmisw}), reduces to
\begin{eqnarray}
\label{eq:emmdis}
	1+\beta_{i\perp}+\beta_{e\perp}-\frac{\beta_{e\perp}(1+\xi_{e}Z_{e})}{2(1+a_{e})}=0,
\end{eqnarray}
with the polarization given by
\begin{eqnarray}
\label{eq:emmpol}
	\frac{\Phi_{\parallel}}{B_{\parallel}}=\frac{\tau}{1+a_{e}}\gg 1.
\end{eqnarray}
Therefore,  unlike its ion-driven counterpart, the electron mirror mode could be dominated by the scalar parallel potential $\Phi_{\parallel}$.

%While it is difficult to prove that $\Omega=0$ is the unique route to cross the real axis, we remark here that Eq. (\ref{eq:critical}) is indeed of value for the analysis of linear instability.

\section{Numerical Results}
\label{sec:nov:07:19:46}
To analyse the linear wave properties numerically, we solve Eq. (\ref{eq:char}) by using a new eigenvalue-solver \citep{chen18,chen2102,chen21}, which can locate all of the eigenvalues in a closed complex frequency domain.
Specifically, the system is controlled by six parameters: $b_{i},a_{i},\beta_{i\perp},\tau,a_{e}$ and the mass ratio $m_{i}/m_{e}$. 

\subsection{Isotropic plasma regime}
As a reference case, we first investigate the isotropic case; i.e., $a_{s}=0$.
As can be anticipated from the instability conditions, Eqs.(\ref{eq:firehose}) and (\ref{eq:ms}), the system is stable in this case.

\begin{figure}[!htp]
\vspace{-0.3cm}
\setlength{\belowcaptionskip}{-0.1cm}
\centering
\includegraphics[scale=0.40]{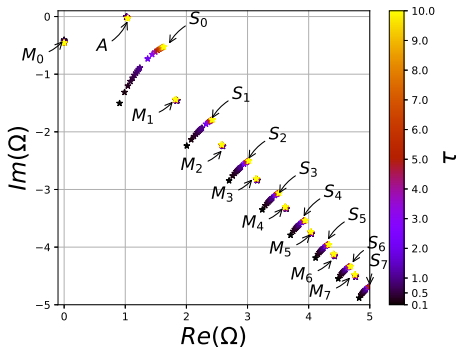}
	\caption{(Color online) Normalized complex frequency $\Omega$ vs $\tau$  in the region $Im(\Omega)\ge -5$ and $0\le Re(\Omega)\le 5$, for $\beta_{i\perp}=1$, $b_{i}=0.1$, $a_{i}=a_{e}=0$ and $m_{i}/m_{e}=1836$.}
\label{eps:iso_tau_eig}
\end{figure}

The complex normal mode frequency as a function of $\tau$ is shown in Fig. (\ref{eps:iso_tau_eig}) for  $\beta_{i\perp}=1$, $b_{i}=0.1$, $a_{i}=a_{e}=0$ and $m_{i}/m_{e}=1836$.
Depending on the  behaviour of wave frequency, three different branches can be identified here: the ISW branch (denoted by $S_{j}$) is sensitive to $\tau$ variation, and the $\textrm{Im}(\Omega)$ becomes less damped with $\tau$; the long-wavelength KAW (denoted by $A$) has $\Omega\simeq 1$; and the mirror branch (denoted by $M_{j}$).We also note that $\tau$ has  weak stabilizing effect on the KAW and MM.
Furthermore, it is also worthwhile mentioning that when $|\xi_{i}|$ is sufficiently large, one has $1+\xi_{i}Z(\xi_{i})\simeq 0$  except for the $\textrm{Im}(\Omega)\simeq-\textrm{Re}(\Omega)$ region \citep{fried}. Therefore, as shown in Fig. (\ref{eps:iso_tau_eig}), the strongly ion Landau damped higher eigenstates of ISW and MM normal modes with large $|\xi_{i}|$ cluster around the $\textrm{Im}(\Omega)\simeq-\textrm{Re}(\Omega)$ region.
%\end{linenumbers}

\begin{figure*}[!htp]
\centering
	\subfloat[\small{MM}]{
\label{eps:iso_tau_pol_mm}
\begin{minipage}[t]{0.32\textwidth}
\centering
\includegraphics[scale=0.30]{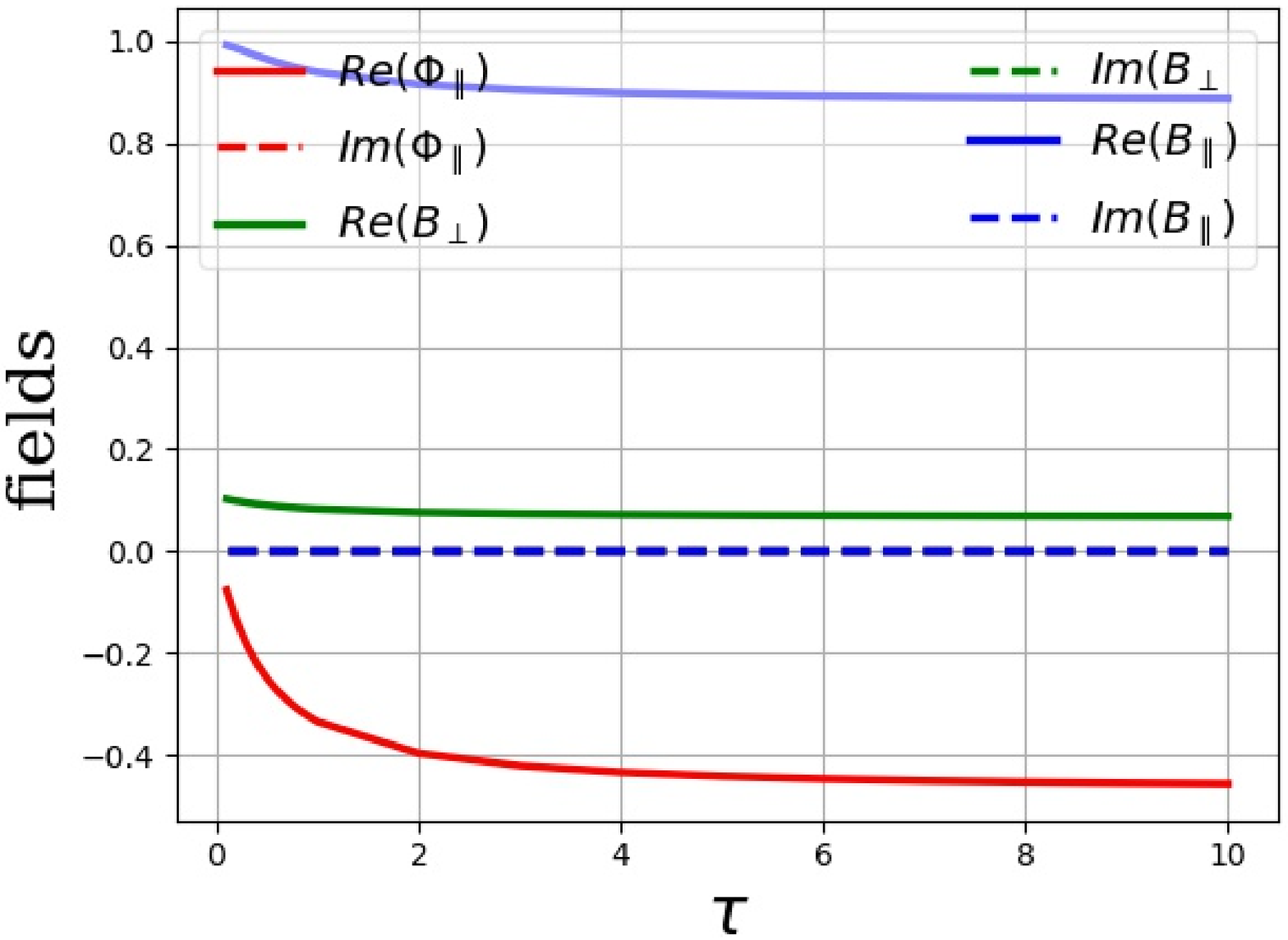}
\end{minipage}
}
	\subfloat[\small{ISW}]{
\label{eps:iso_tau_pol_is}
\begin{minipage}[t]{0.32\textwidth}
\centering
\includegraphics[scale=0.30]{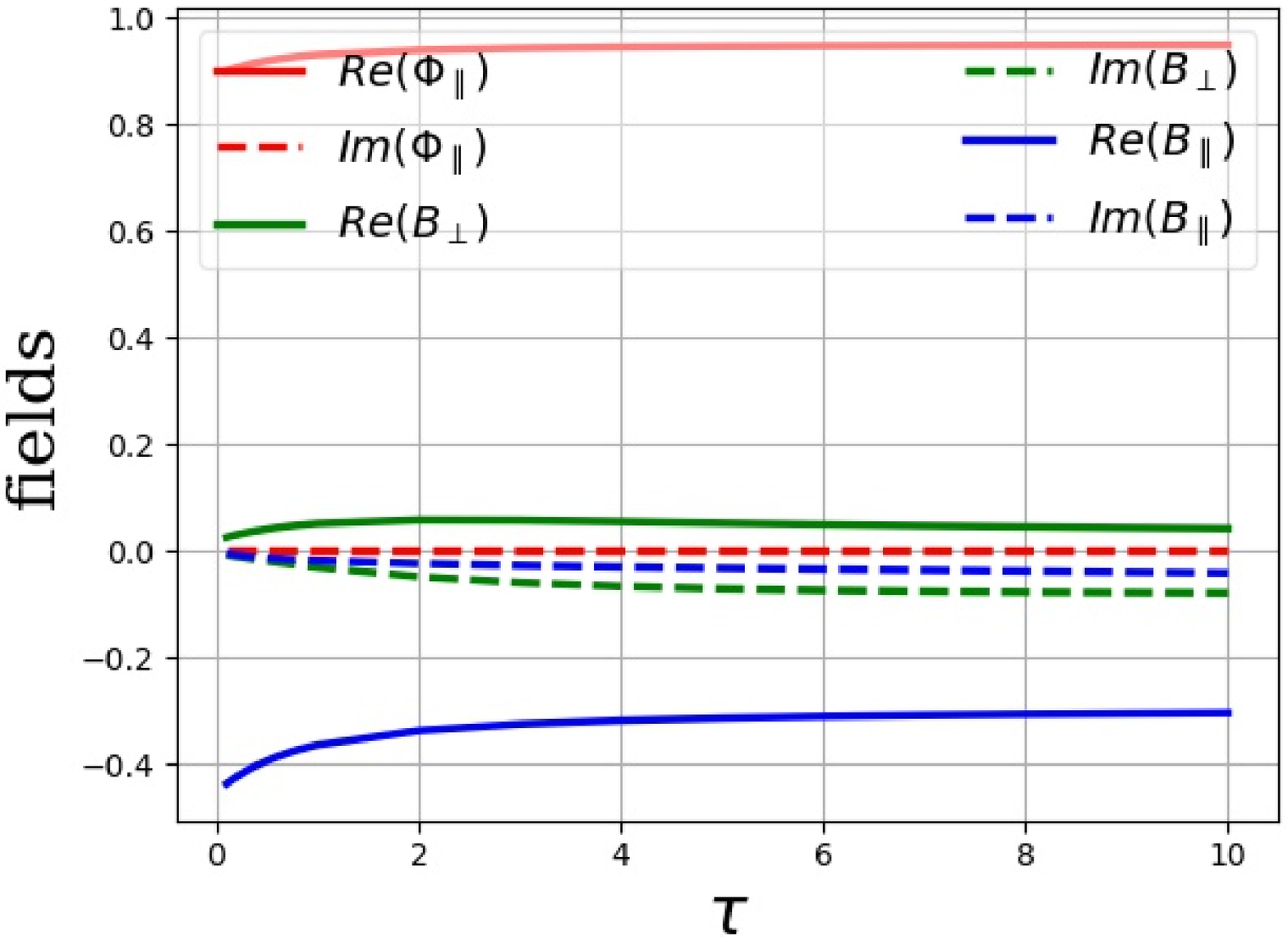}
\end{minipage}
}
	\subfloat[\small{KAW}]{
\label{eps:iso_tau_pol_al}
\begin{minipage}[t]{0.32\textwidth}
\centering
\includegraphics[scale=0.30]{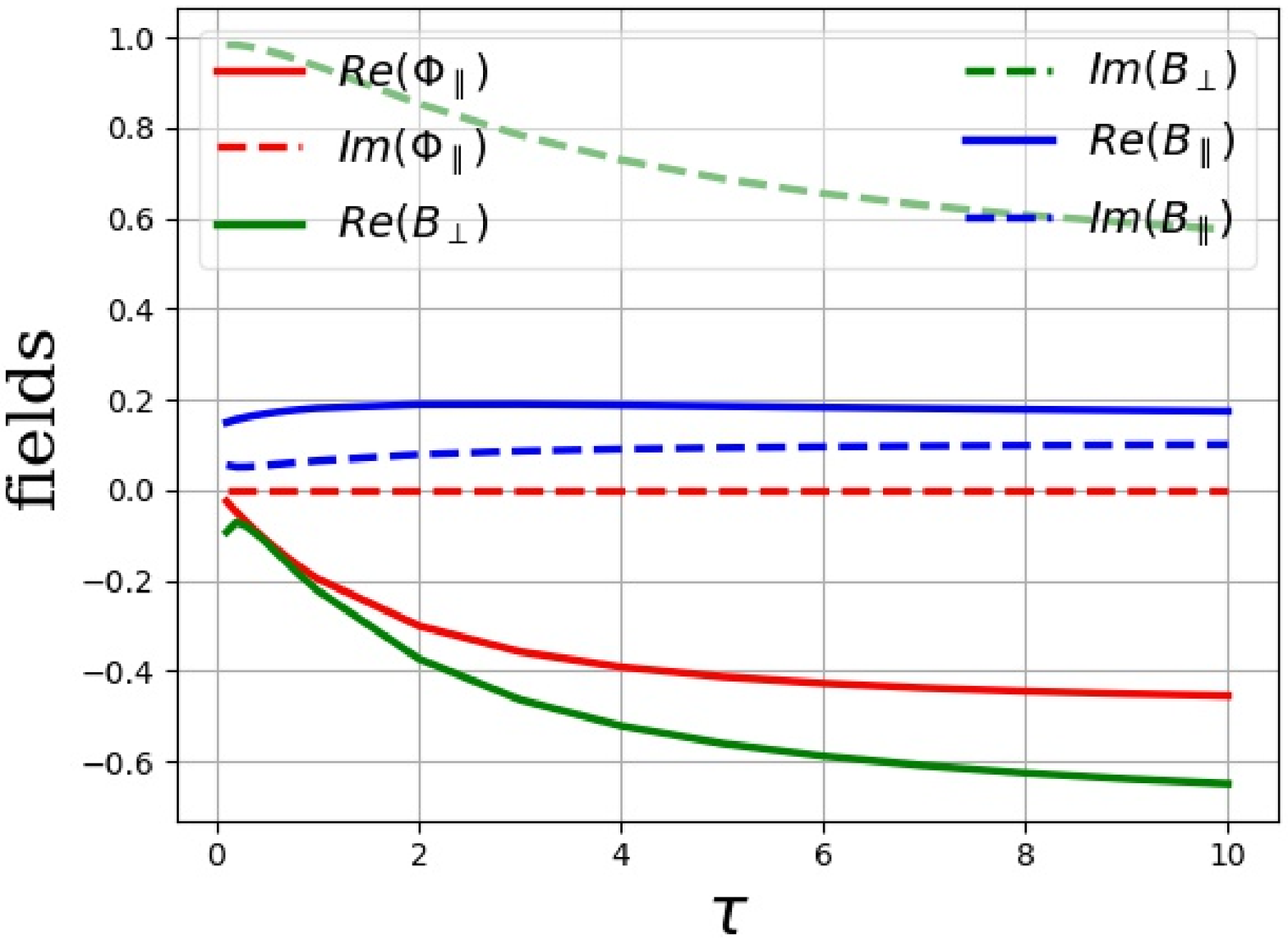}
\end{minipage}
}
\caption{(Color online) Plots of wave polarizations vs $\tau$. The other parameters are the same as in Fig. (\ref{eps:iso_tau_eig}).}
\label{eps:pol_tau}
\end{figure*}

%\begin{linenumbers}
Since the wave polarization plays crucial roles in the particle acceleration, heating and cross-field transport \citep{chen99}, we also investigate how wave polarization varies with parameters.
In the present work, we only consider the polarization of the KAW, and the least damped MM and ISW, i.e., $M_{0}$ and $S_{0}$ in Fig. (\ref{eps:iso_tau_eig}).

Figure (\ref{eps:pol_tau}a) shows the wave polarization of the MM with increasing $\tau$. It is found that the magnetic compressibility $B_{\parallel}$ decreases slightly, whereas the scalar potential $\Phi_{\parallel}$ shows a significant increase.
The perpendicular magnetic field, normalized as $B_{\perp}\equiv\delta B_{\perp}/B$, is insensitive to $\tau$.
The polarization of ion-sound wave, as illustrated in Fig.(\ref{eps:pol_tau}b), on the other hand, has a weak dependence on $\tau$.
By comparing Fig. (\ref{eps:pol_tau}a) and (\ref{eps:pol_tau}b), we note that the MM and ISW are dominated, respectively, by the $B_{\parallel}$ and $\Phi_{\parallel}$, and the fields $B_{\parallel}$ and $\Phi_{\parallel}$ are opposite in phase in both MM and ISW.
This has the implication that the mirror force and parallel electric force  are in the same direction for electrons and opposite for ions.
As depicted in Fig. (\ref{eps:pol_tau}c), $B_{\perp}$ dominates in KAW as expected, $B_{\parallel}$ is nearly independent of $\tau$.
$\Phi_{\parallel}$, meanwhile, increases with the increasing $\tau$.
%\end{linenumbers}
%while the magnitude of perpendicular magnetic field $B_{\perp}\equiv\delta B_{\perp}/B=(1+a_{i})\Psi\sqrt{0.5\beta_{i\perp}}/\Omega$ is less sensitive to the $\tau$ variation.

\begin{figure}[!htp]
\vspace{-0.3cm}
\setlength{\belowcaptionskip}{-0.1cm}
\centering
\includegraphics[scale=0.40]{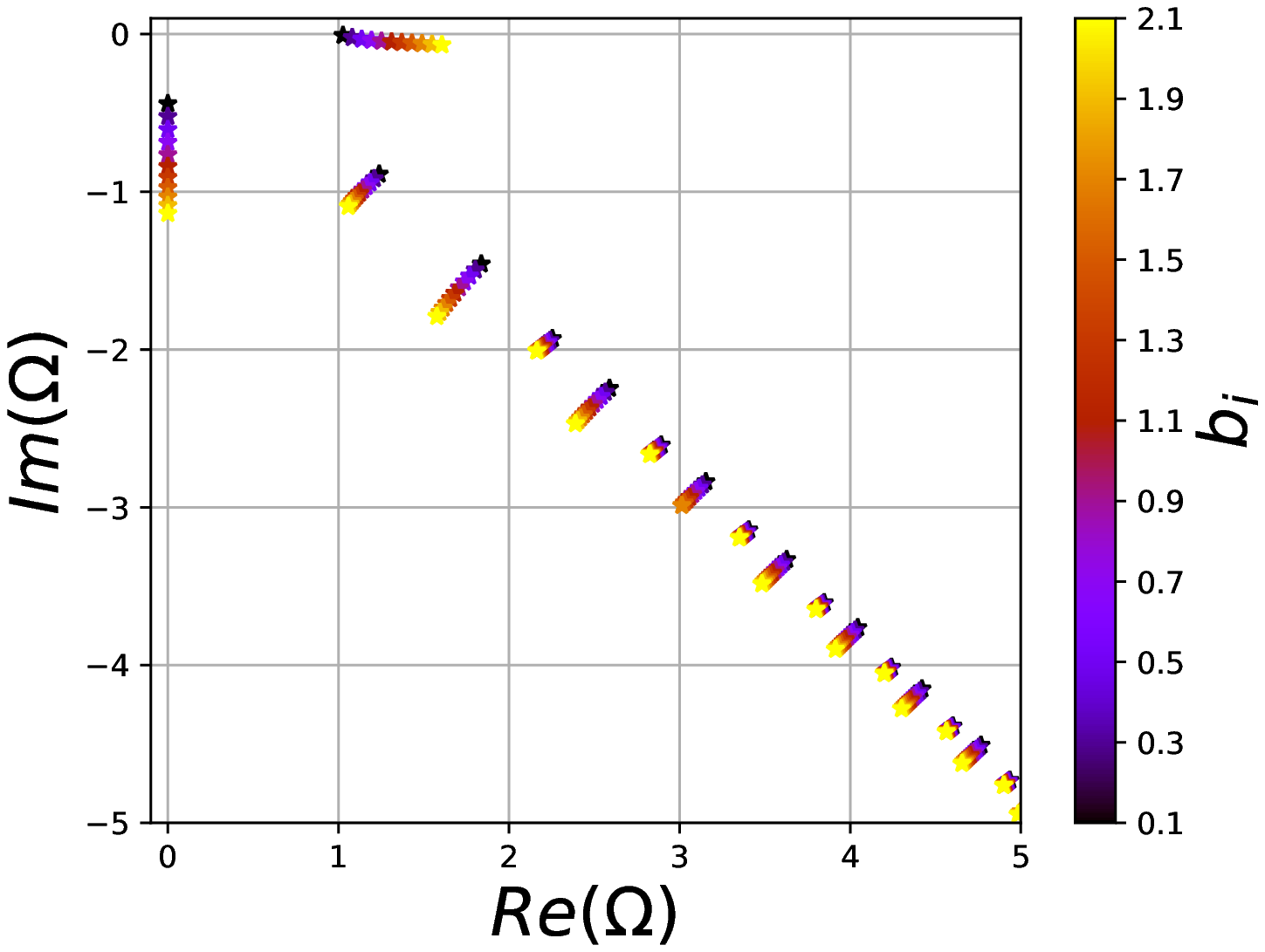}
	\caption{(Color online) Normalized complex frequency $\Omega$ vs $b_{i}$  in the region $Im(\Omega)\ge -5$ and $0\le Re(\Omega)\le 5$, for $\beta_{i\perp}=1$, $\tau=1$, $a_{i}=a_{e}=0$ and $m_{i}/m_{e}=1836$.}
\label{eps:iso_bi_eig}
\end{figure}
\begin{figure*}[!htp]
\centering
	\subfloat[\small{MM}]{
\label{eps:iso_bi_pol_mm}
\begin{minipage}[t]{0.32\textwidth}
\centering
\includegraphics[scale=0.30]{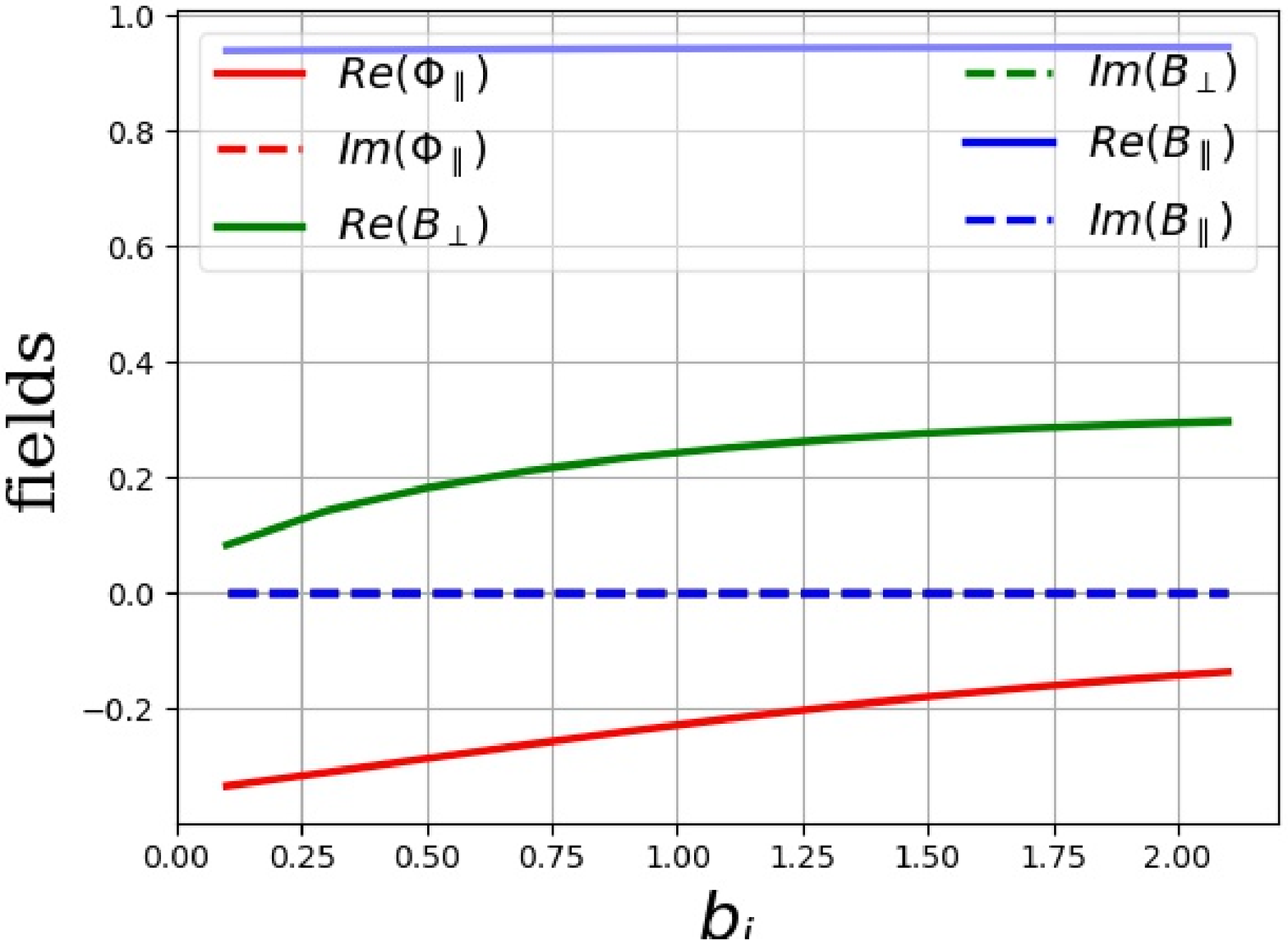}
\end{minipage}
}
	\subfloat[\small{ISW}]{
\label{eps:iso_bi_pol_is}
\begin{minipage}[t]{0.32\textwidth}
\centering
\includegraphics[scale=0.30]{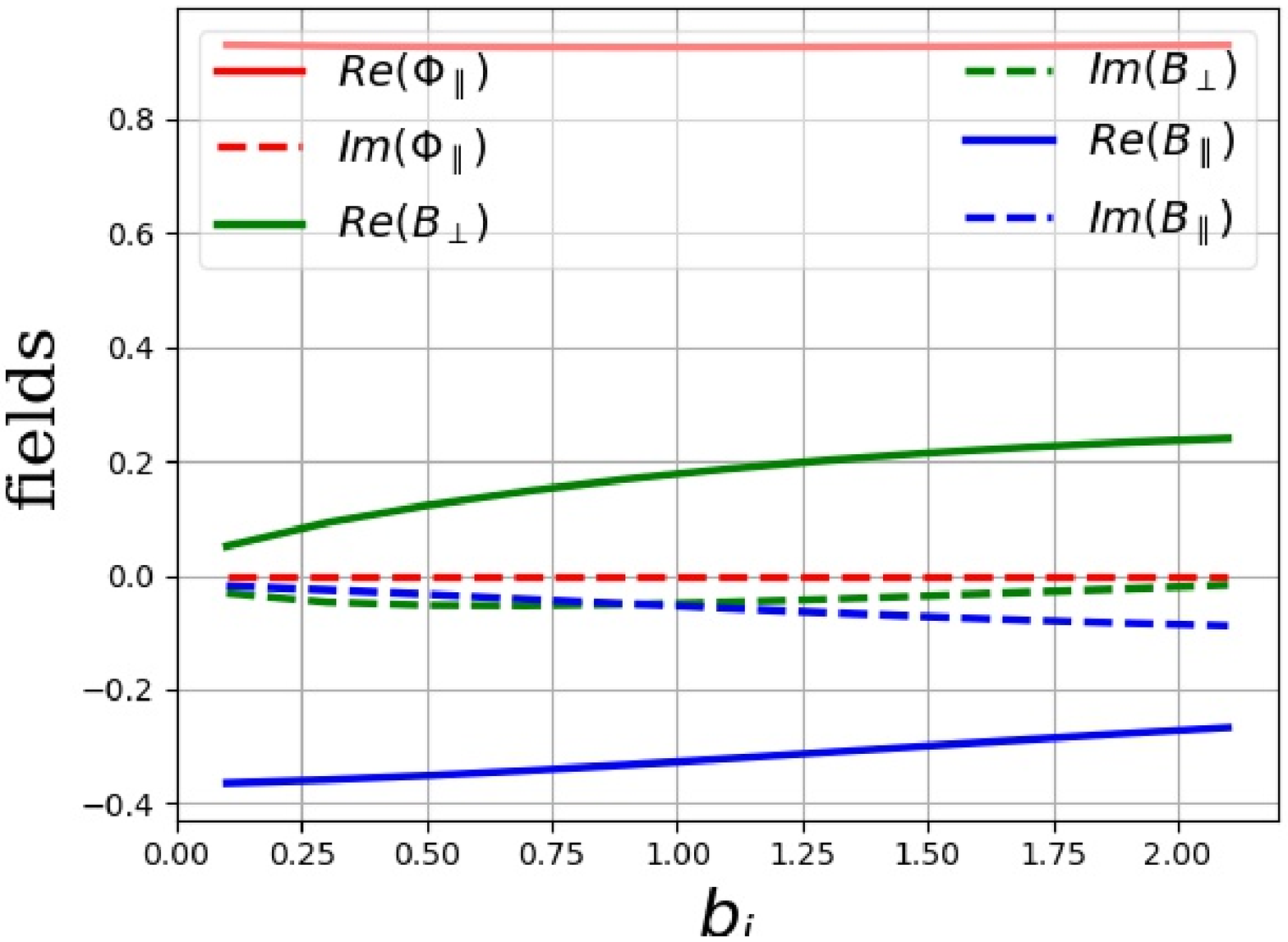}
\end{minipage}
}
	\subfloat[\small{KAW}]{
\label{eps:iso_bi_pol_al}
\begin{minipage}[t]{0.32\textwidth}
\centering
\includegraphics[scale=0.30]{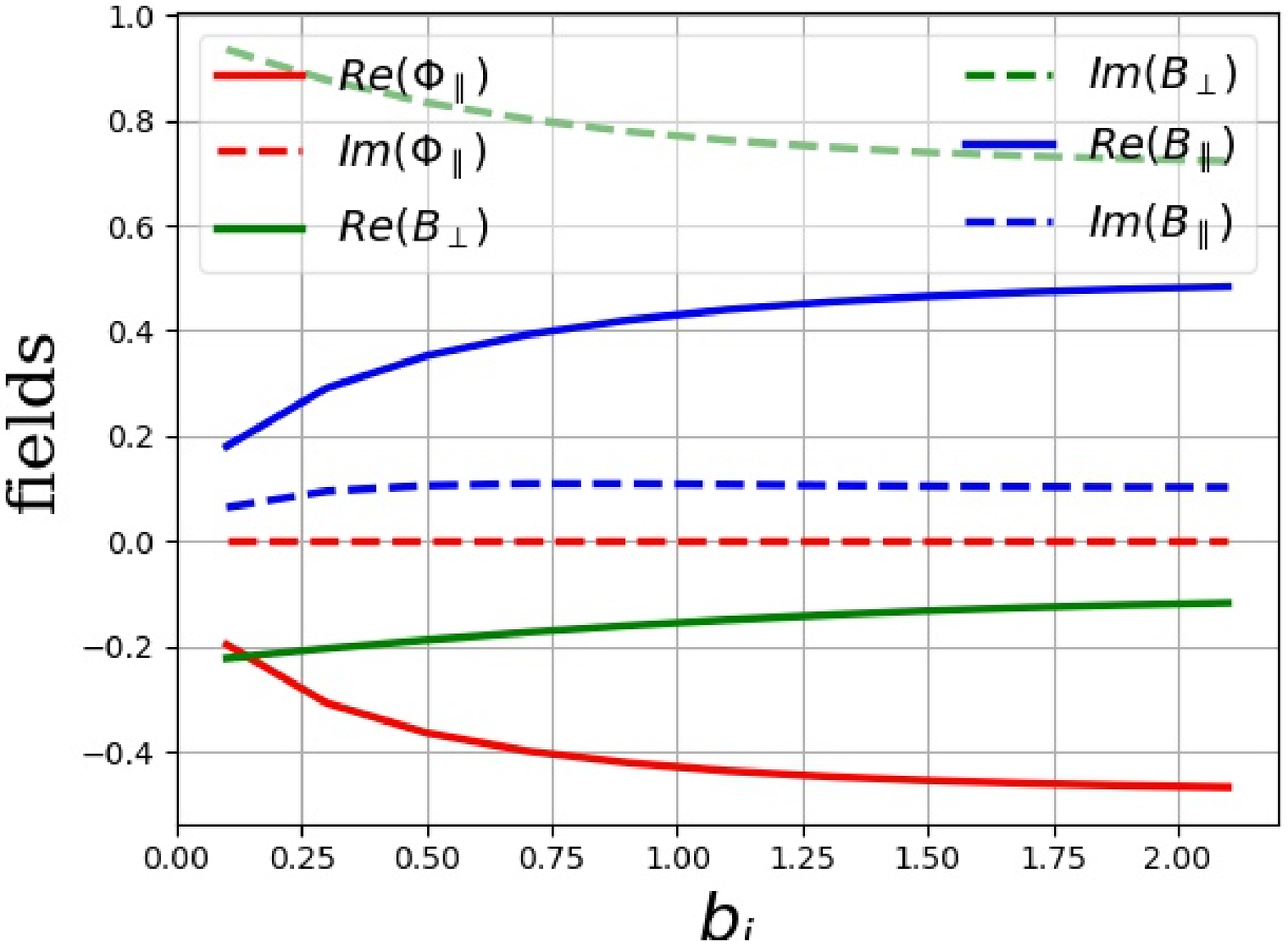}
\end{minipage}
}
\caption{(Color online) Plots of wave polarization vs $b_{i}$. The other parameters are the same as in Fig. (\ref{eps:iso_bi_eig}). }
\label{eps:pol_bi}
\end{figure*}

%\begin{linenumbers}
The complex normal mode frequency as  a function of $b_{i}$ is given in Fig.(\ref{eps:iso_bi_eig}). It is clear that FILR has a stabilizing effect on all of the low-frequency waves. The MM branch is more sensitive to $b_{i}$ than the ISW branch. As  $b_{i}$ is increased, the real frequency of Alfv\'{e}n wave, consistent with KAW dispersion relation Eq. (\ref{eq:lowbetadisper1}), increases significantly, while the real frequency of MM-ISW branch decreases.
Shown in Fig. (\ref{eps:pol_bi}) are the results of a $b_{i}$ scan for wave polarizations.
Notably, as $b_{i}$ increases, both $B_{\parallel}$ and $\Phi_{\parallel}$ of KAW increase significantly and remain opposite in phase. 
%\end{linenumbers}

\begin{figure}[!htp]
\vspace{-0.3cm}
\setlength{\belowcaptionskip}{-0.1cm}
\centering
\includegraphics[scale=0.40]{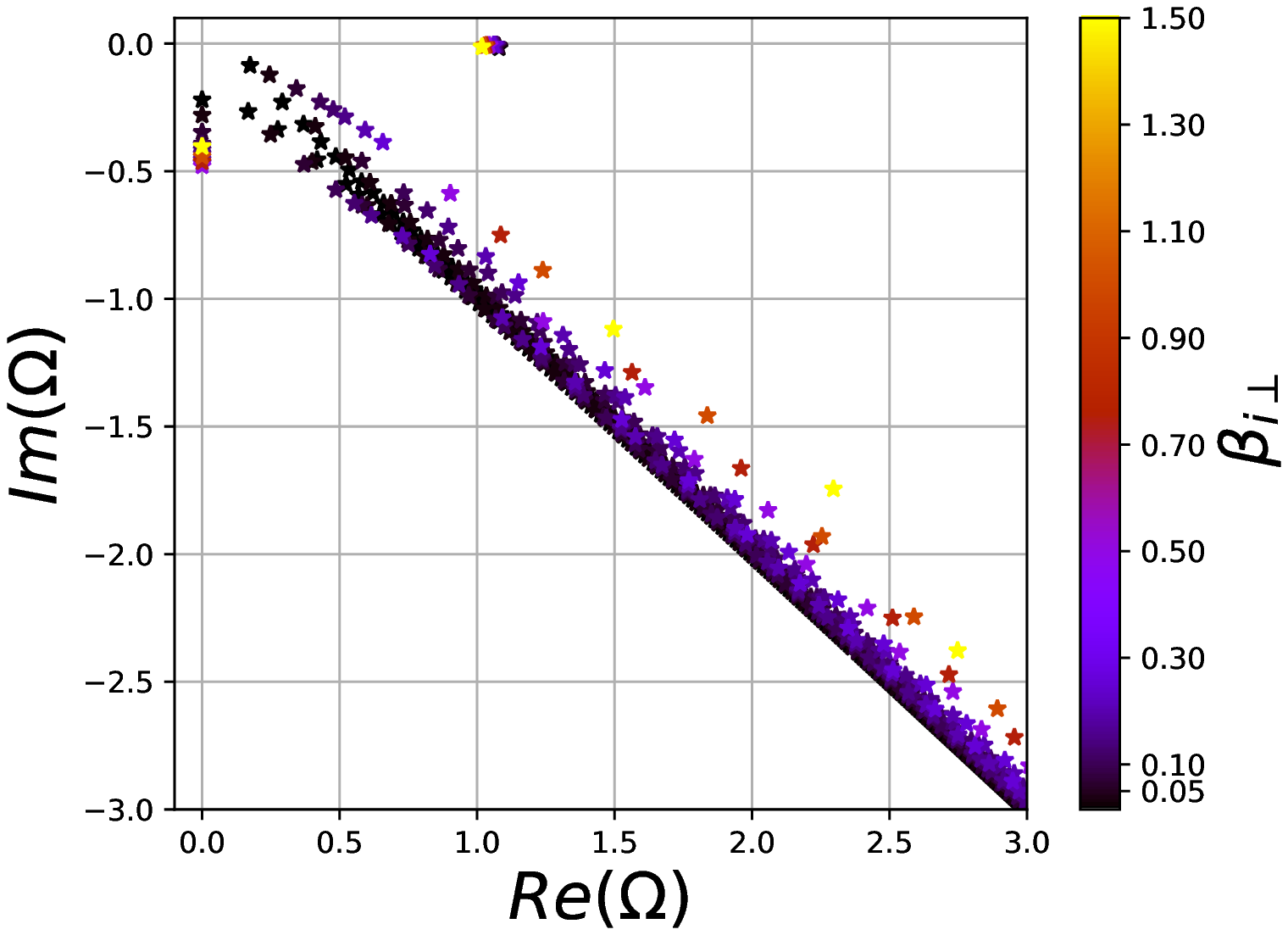}
	\caption{(Color online) Normalized complex frequency $\Omega$ vs $\beta_{i\perp}$  in the region $Im(\Omega)\ge -3$ and $0\le Re(\Omega)\le 3$, for $b_{i}=0.1$, $\tau=1$, $a_{i}=a_{e}=0$ and $m_{i}/m_{e}=1836$.}
\label{eps:iso_betai_eig}
\end{figure}
\begin{figure*}[!htp]
\centering
	\subfloat[\small{MM}]{
\label{eps:iso_betai_pol_mm}
\begin{minipage}[t]{0.32\textwidth}
\centering
\includegraphics[scale=0.30]{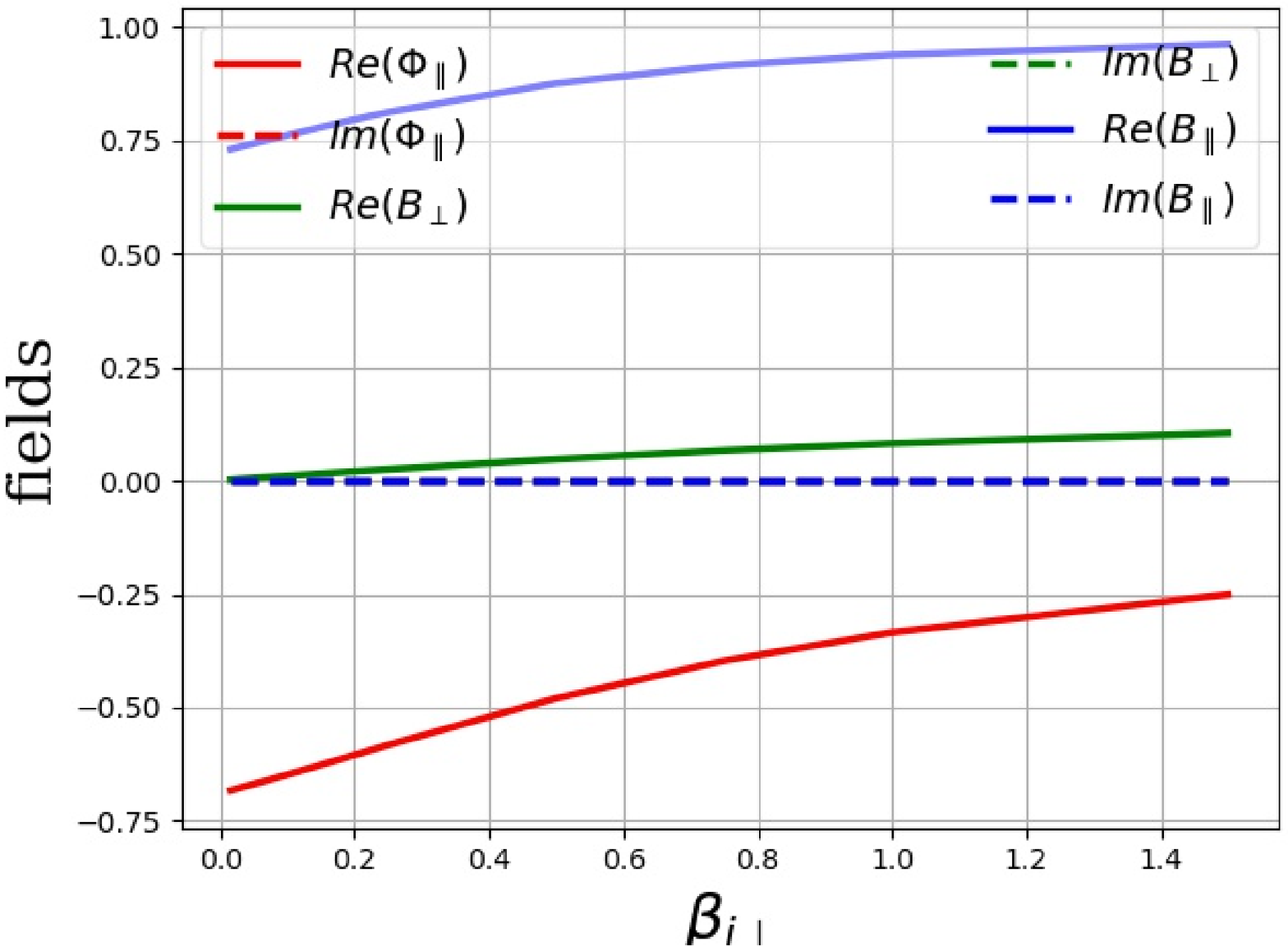}
\end{minipage}
}
	\subfloat[\small{ISW}]{
\label{eps:iso_betai_pol_is}
\begin{minipage}[t]{0.32\textwidth}
\centering
\includegraphics[scale=0.30]{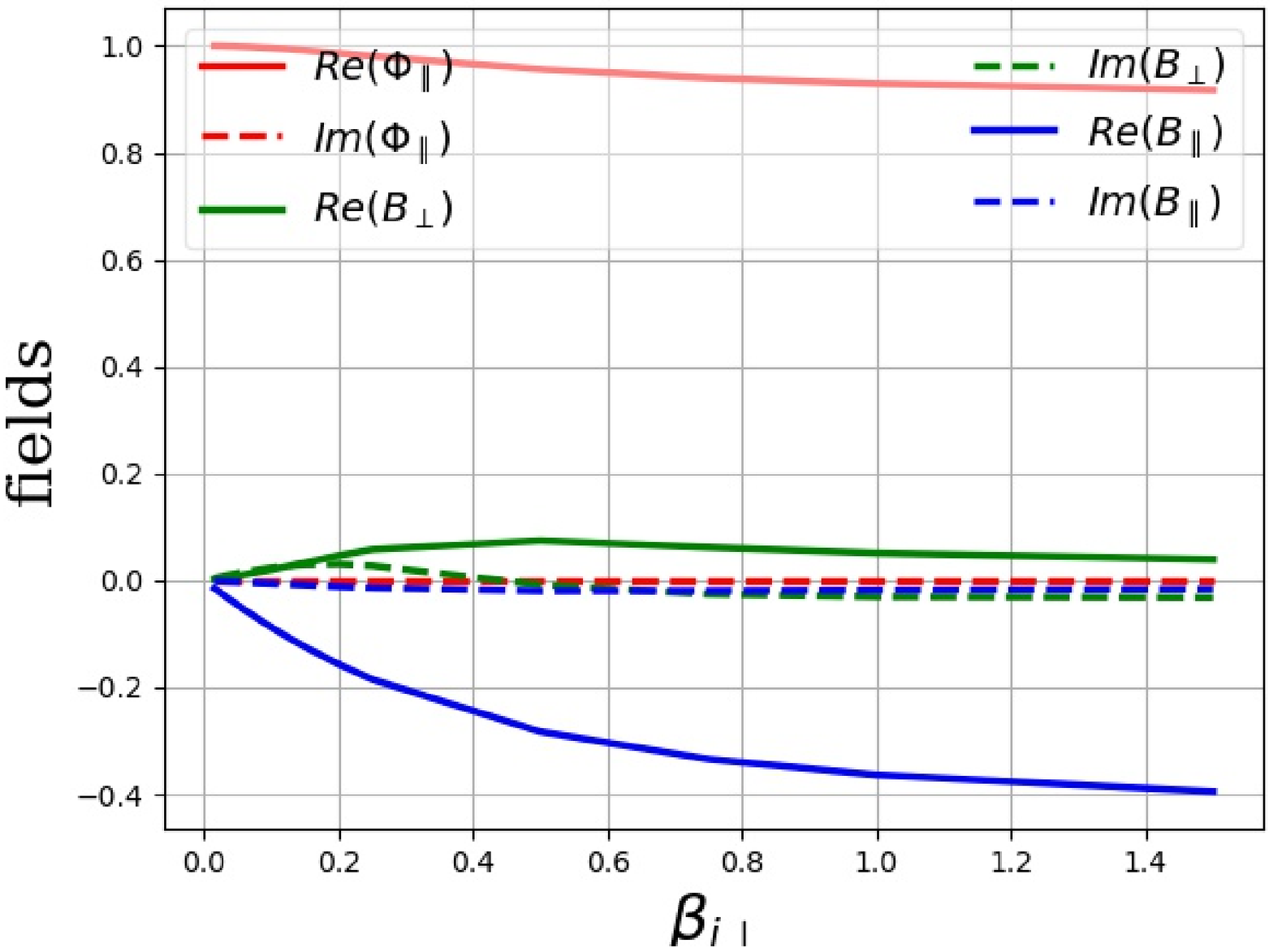}
\end{minipage}
}
	\subfloat[\small{KAW}]{
\label{eps:iso_betai_pol_al}
\begin{minipage}[t]{0.32\textwidth}
\centering
\includegraphics[scale=0.30]{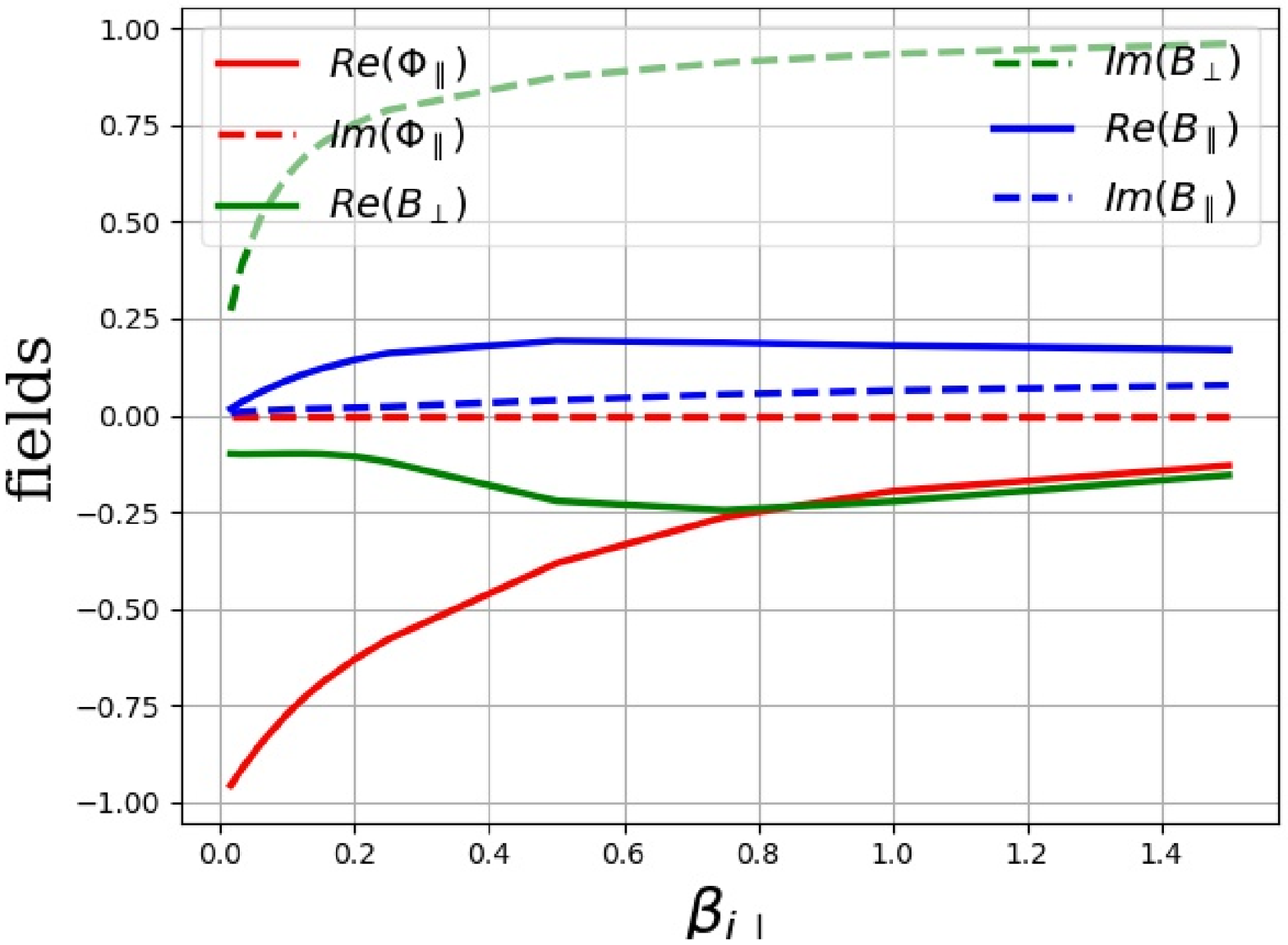}
\end{minipage}
}
\caption{(Color online) Plots of wave polarization vs $\beta_{i\perp}$. The other parameters are the same as in Fig. (\ref{eps:iso_betai_eig}).}
\label{eps:pol_betai}
\end{figure*}

%\begin{linenumbers}
Figure (\ref{eps:iso_betai_eig}) shows the dependence of the complex normal mode frequencies on $\beta_{i\perp}$.
Since $\beta_{i\perp}$ is a basic parameter determining the magnitude of $v^{2}_{ti\parallel}/v^{2}_{A}=(1+a_{i})\beta_{i\perp}$,  then, in the $\beta_{i\perp}\ll 1$ limit, we find, again, that those modes with $|\xi_{i}|\gg 1$ and clustering around $\textrm{Im}(\Omega)\simeq-\textrm{Re}(\Omega)$, correspond to heavily ion Landau damped higher eigenstates of MM and ISW normal modes.
Interestingly, $\beta_{i\perp}$ has a further stabilizing effect on the MM in this isotropic limit.
$\beta_{i\perp}$ can also strongly stabilize ISWs, however, here this effect can be mainly attributed to the coupling to MMs, as will be shown below in Fig. (\ref{eps:pol_betai}b).
The corresponding wave polarizations are depicted in Fig. (\ref{eps:pol_betai}).
As $\beta_{i\perp}$ increases, $|\Phi_{\parallel}/B_{\parallel}|$ decreases significantly for the MM. 
Also, $|\Phi_{\parallel}/B_{\perp}|$ decreases similarly for the KAW.
Meanwhile,  $|B_{\parallel}/|\Phi_{\parallel}|$ increases significantly for the  ISW, indicating increasing coupling to MM.
%\end{linenumbers}

\subsection{Mirror instability regime}

\begin{figure}[!htp]
\vspace{-0.3cm}
\setlength{\belowcaptionskip}{-0.1cm}
\centering
\includegraphics[scale=0.40]{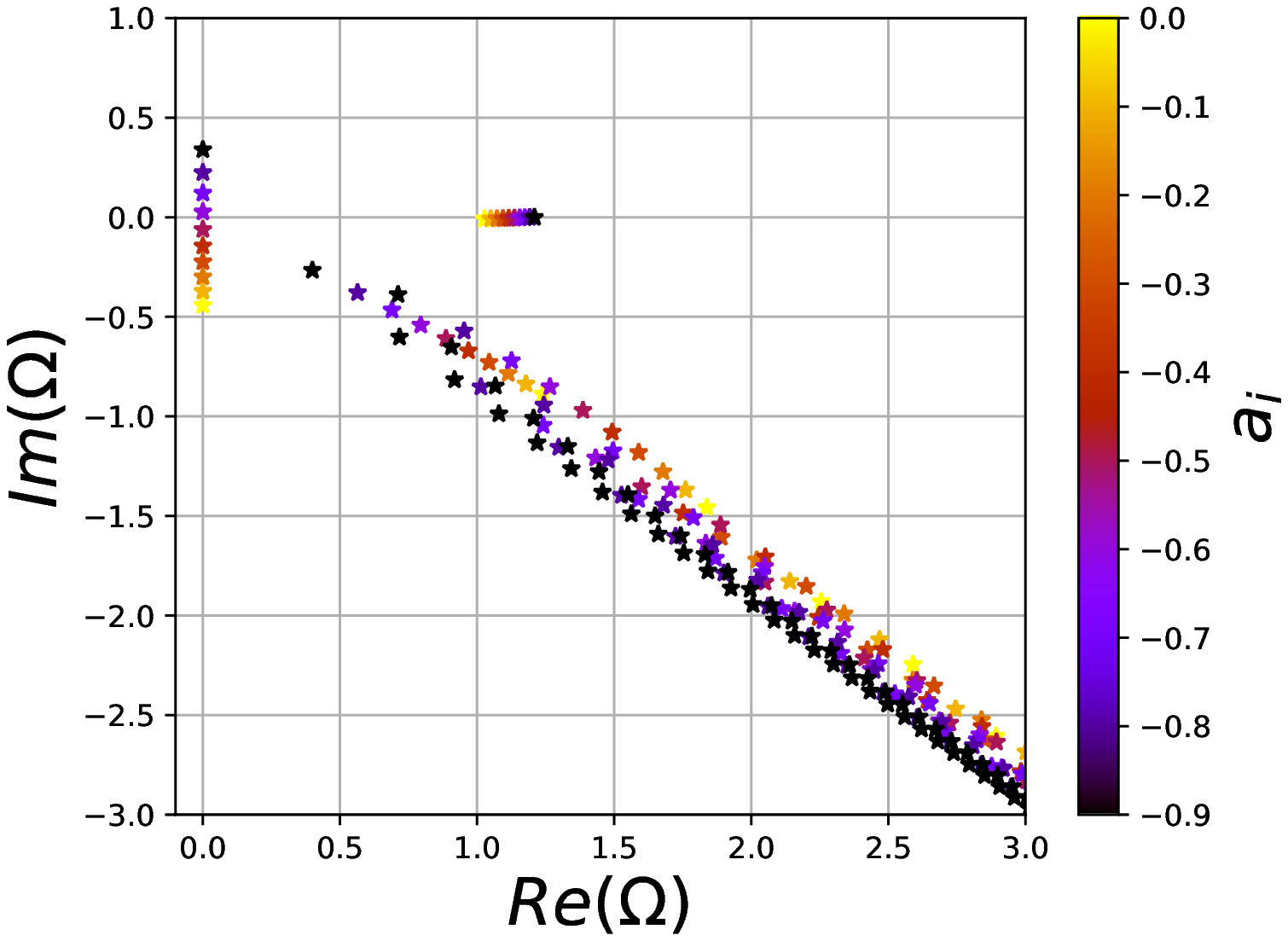}
	\caption{(Color online) Normalized complex frequency $\Omega$ vs $a_{i}$  in the region $-3\le Im(\Omega)\le 1$ and $0\le Re(\Omega)\le 3$, for $b_{i}=0.1$, $\beta_{i\perp}=1$, $\tau=1$, $a_{e}=0$ and $m_{i}/m_{e}=1836$.}
\label{eps:mm_ai_eig}
\end{figure}
\begin{figure*}[!htp]
\centering
	\subfloat[\small{MM}]{
\label{eps:mm_ai_pol_mm}
\begin{minipage}[t]{0.32\textwidth}
\centering
\includegraphics[scale=0.30]{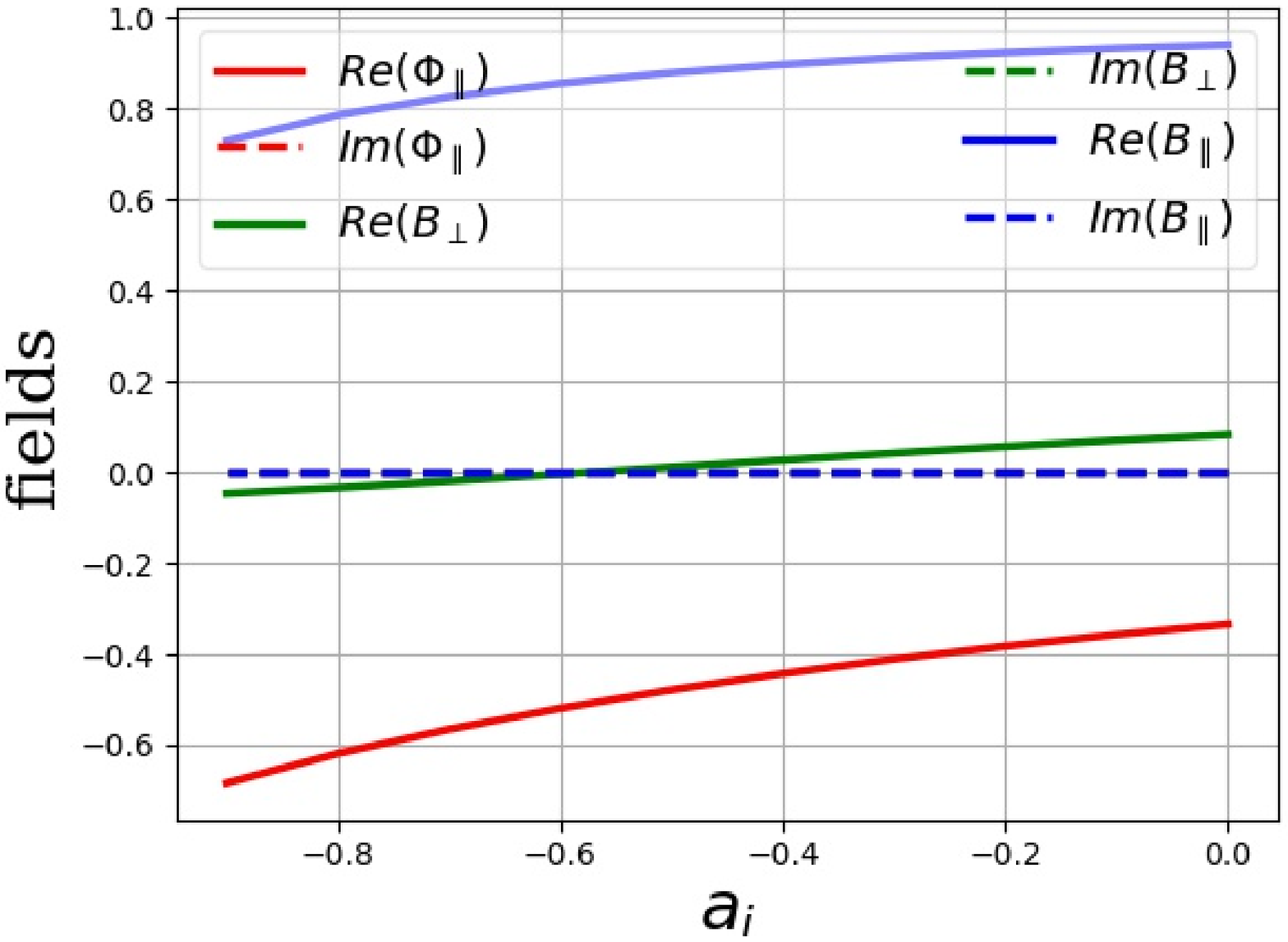}
\end{minipage}
}
	\subfloat[\small{ISW}]{
\label{eps:mm_ai_pol_is}
\begin{minipage}[t]{0.32\textwidth}
\centering
\includegraphics[scale=0.30]{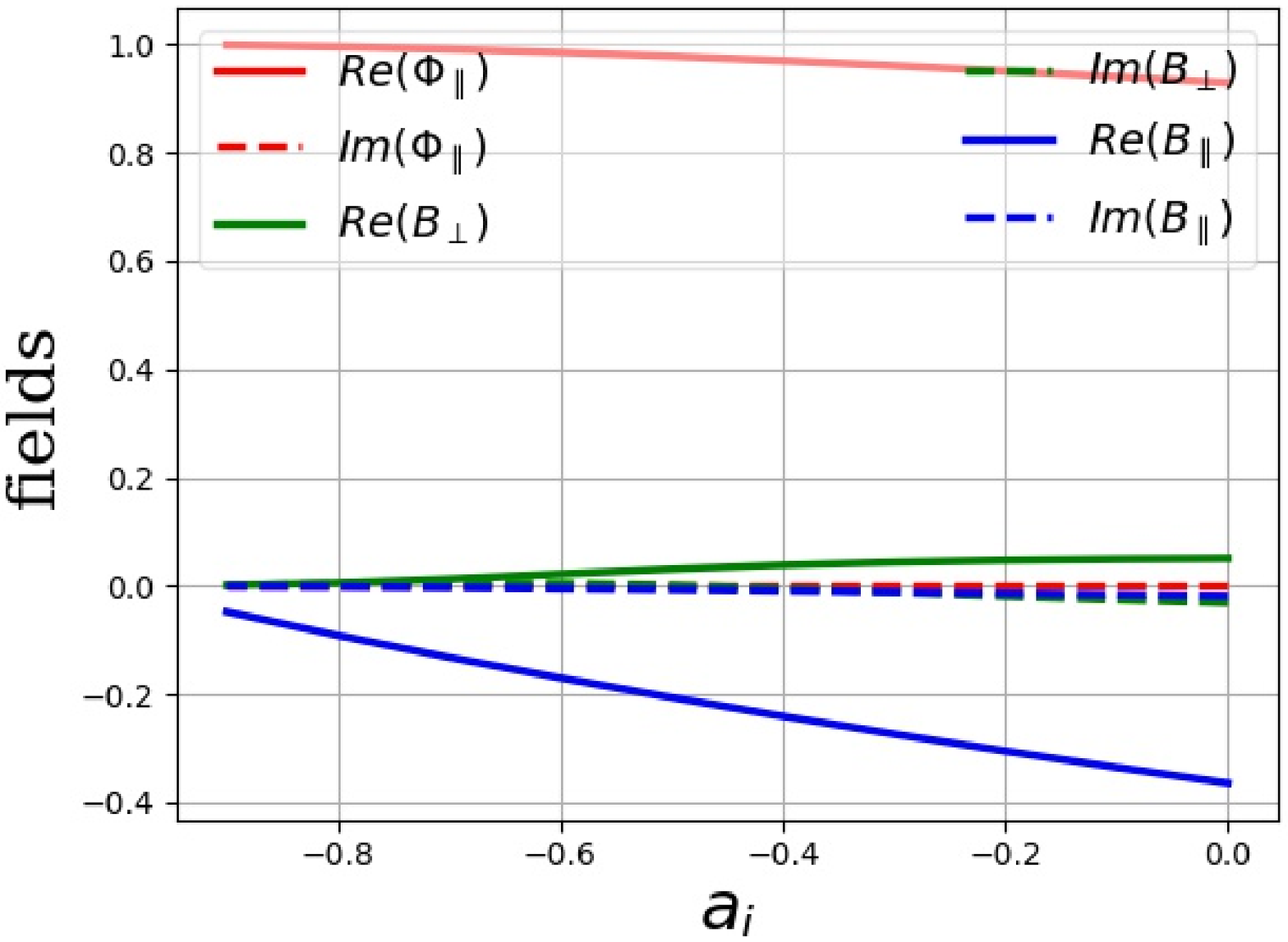}
\end{minipage}
}
	\subfloat[\small{KAW}]{
\label{eps:mm_ai_pol_al}
\begin{minipage}[t]{0.32\textwidth}
\centering
\includegraphics[scale=0.30]{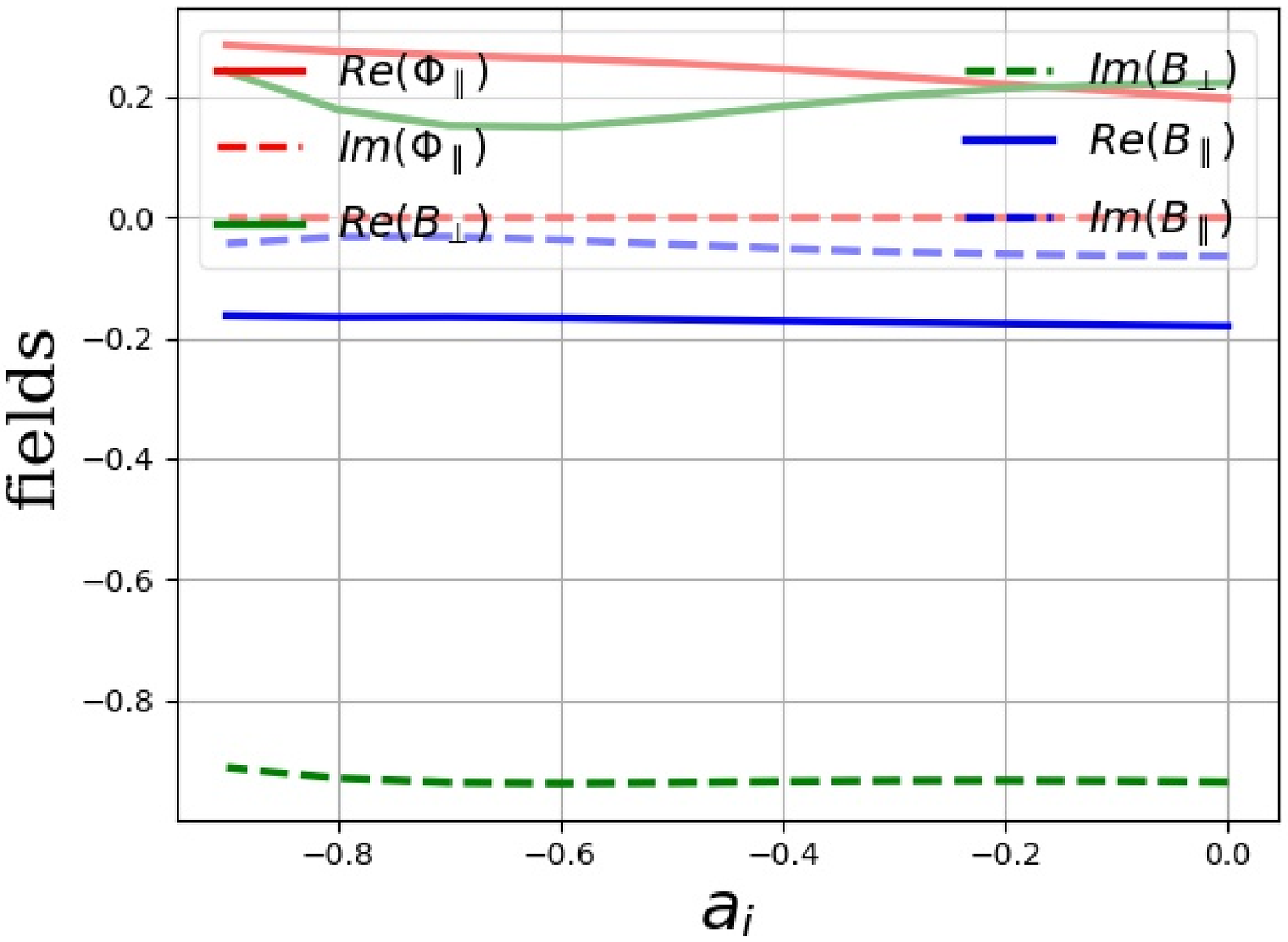}
\end{minipage}
}
	\caption{(Color online) Plots of wave polarization vs $a_{i}$. The other parameters are the same as in Fig. (\ref{eps:mm_ai_eig}).}
\label{eps:mm_pol_ai}
\end{figure*}

%\begin{linenumbers}
Next, we investigate the linear wave properties in the mirror instability regime with $a_{s}<0$ ($T_{s\parallel}<T_{s\perp}$). In order to illustrate the physics more clearly with respect to the isotropic case, we first  explore the ion-driven mirror instability while keeping electron isotropic, i.e., $a_{e}=0$.

Figure (\ref{eps:mm_ai_eig}) shows the normal mode frequency $\Omega$ as a function of $a_{i}$.
It is clear that the mirror instability sets in at sufficiently negative $a_{i}$.
The corresponding wave polarizations are shown in Fig. (\ref{eps:mm_pol_ai}).
While as $a_{i}$ decreases, $|\Phi_{\parallel}/B_{\parallel}|$ increases for the MM (c.f., Fig. (\ref{eps:mm_pol_ai}a)), $|B_{\parallel}/\Phi_{\parallel}|$ decreases significantly for the ISW. Meanwhile, the wave polarization of the KAW shows little variation with $a_{i}$.
%\end{linenumbers}

\begin{figure}[!htp]
\vspace{-0.3cm}
\setlength{\belowcaptionskip}{-0.1cm}
\centering
\includegraphics[scale=0.40]{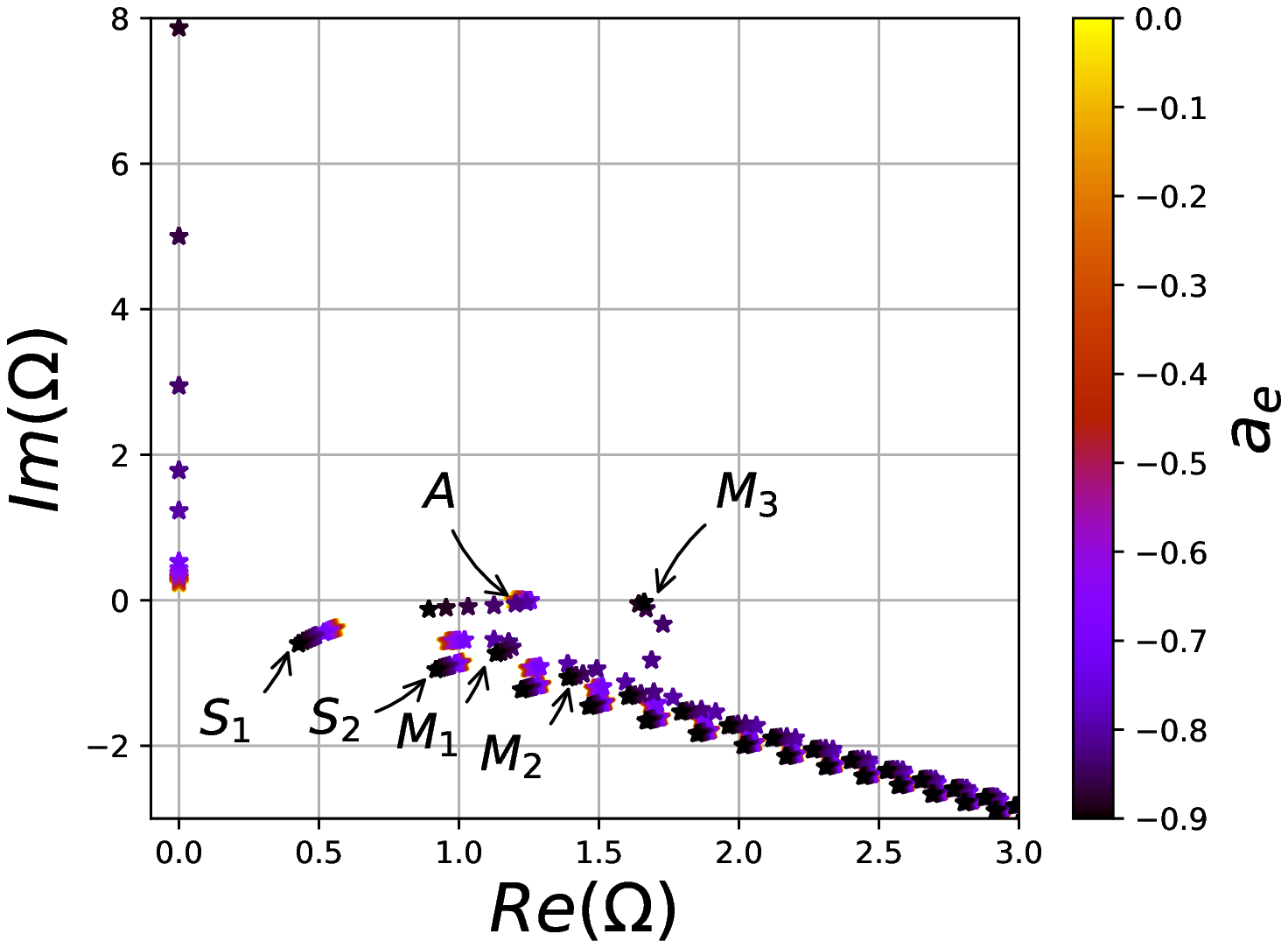}
	\caption{(Color online) Normalized complex frequency $\Omega$ vs $a_{e}$  in the region $-3\le Im(\Omega)\le 8$ and $0\le Re(\Omega)\le 3$, for $b_{i}=0.1$, $\beta_{i\perp}=1$, $\tau=1$, $a_{i}=-0.8$ and $m_{i}/m_{e}=1836$.}
\label{eps:mm_ae_eig}
\end{figure}
\begin{figure*}[!htp]
\centering
	\subfloat[\small{MM ($M_{0}$)}]{
\label{eps:mm_ae_pol_mm}
\begin{minipage}[t]{0.24\textwidth}
\centering
\includegraphics[scale=0.24]{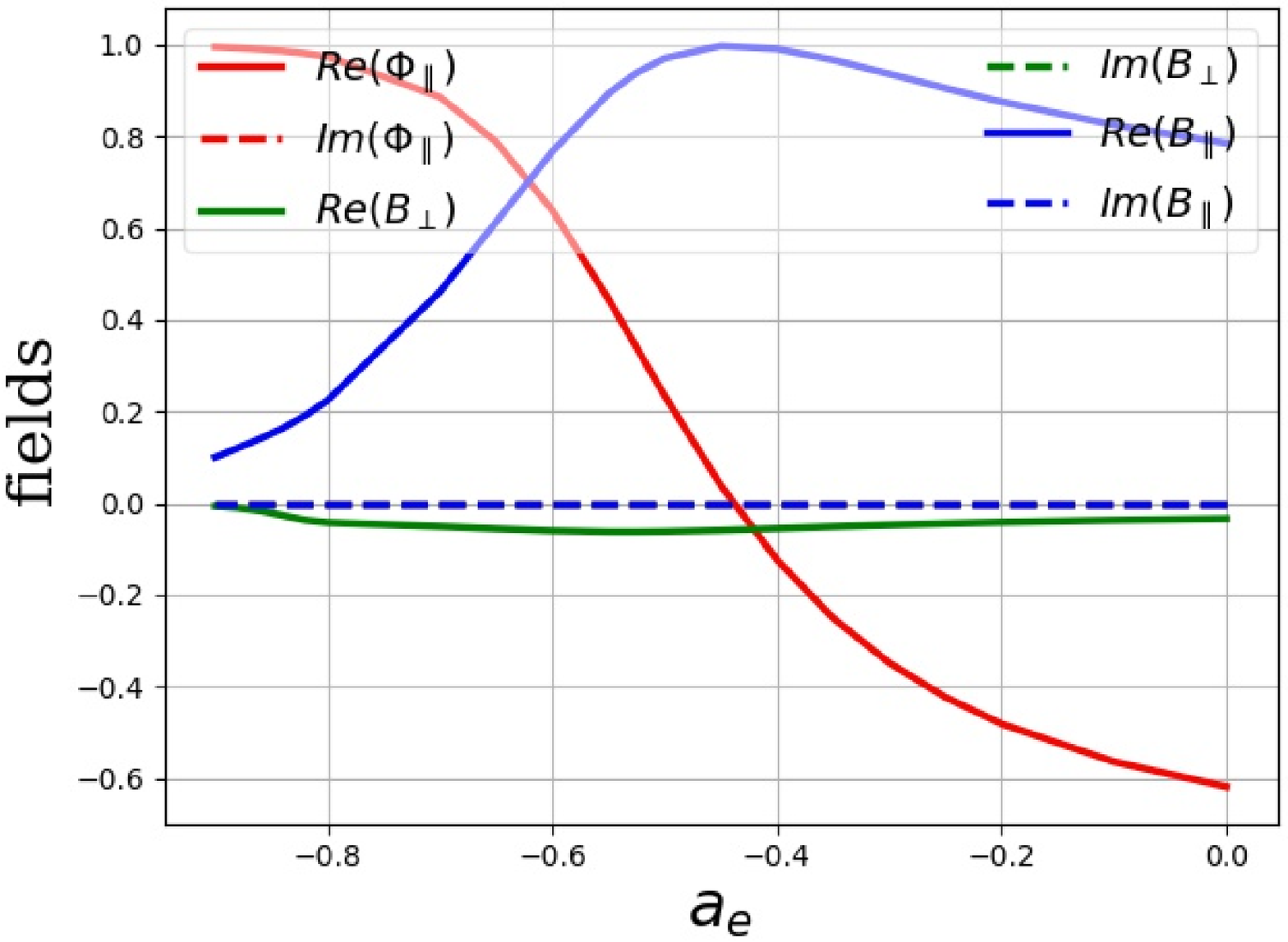}
\end{minipage}
}
	\subfloat[\small{ISW}]{
\label{eps:mm_ae_pol_is}
\begin{minipage}[t]{0.24\textwidth}
\centering
\includegraphics[scale=0.24]{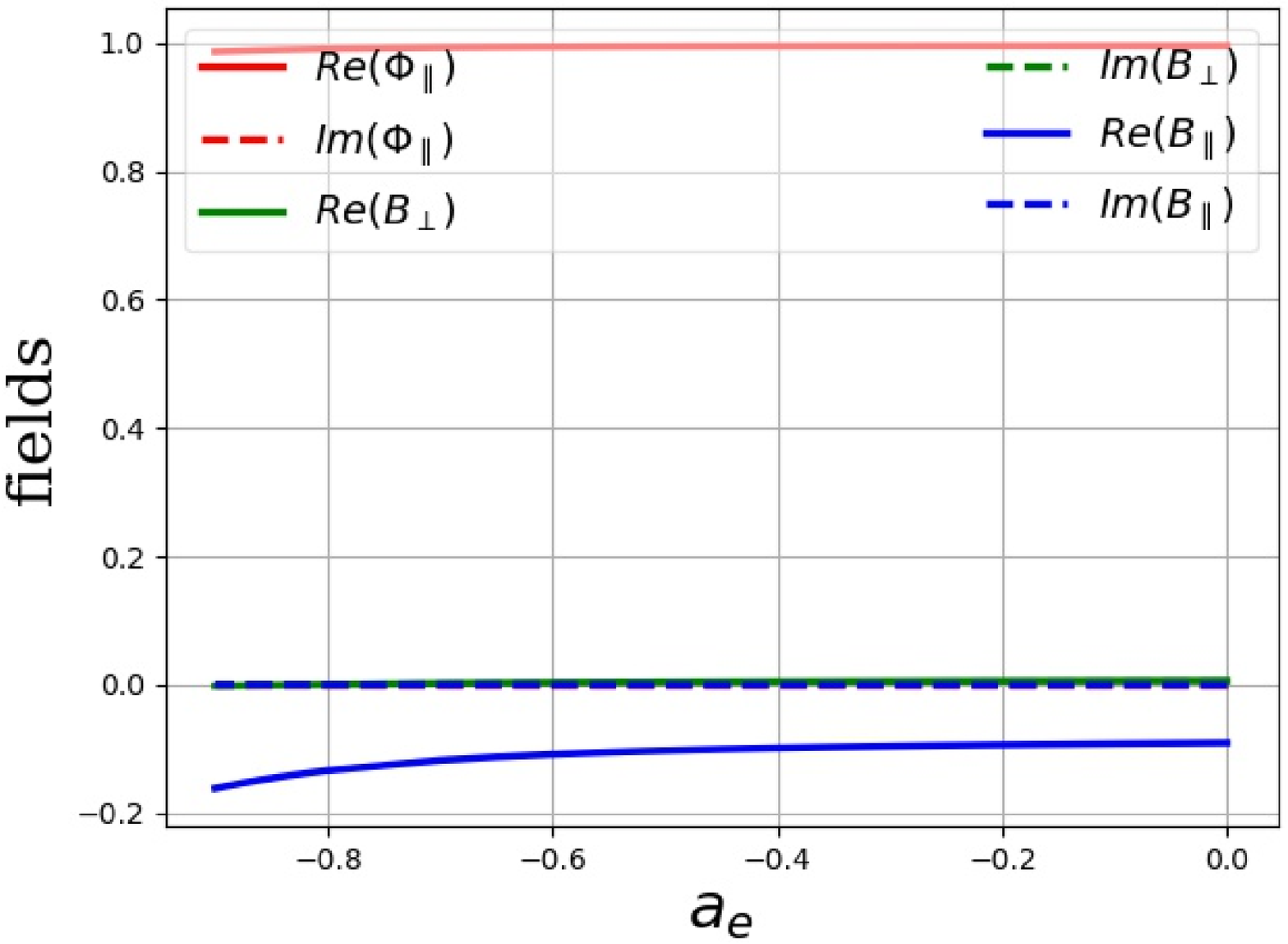}
\end{minipage}
}
	\subfloat[\small{KAW}]{
\label{eps:mm_ae_pol_al}
\begin{minipage}[t]{0.24\textwidth}
\centering
\includegraphics[scale=0.24]{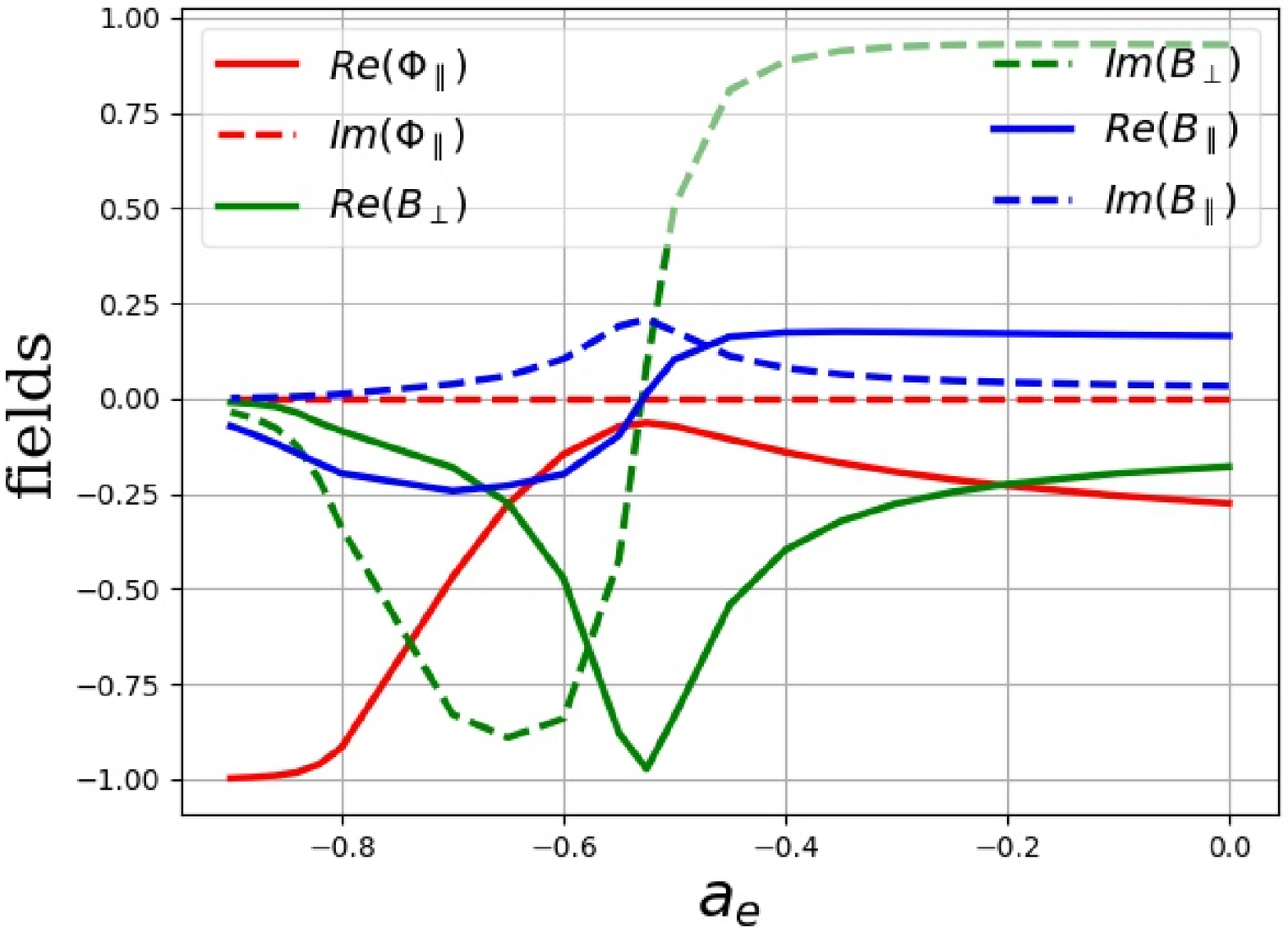}
\end{minipage}
}
	\subfloat[\small{MM ($M_{3}$)}]{
\label{eps:mm_ae_pol_mm3}
\begin{minipage}[t]{0.24\textwidth}
\centering
\includegraphics[scale=0.24]{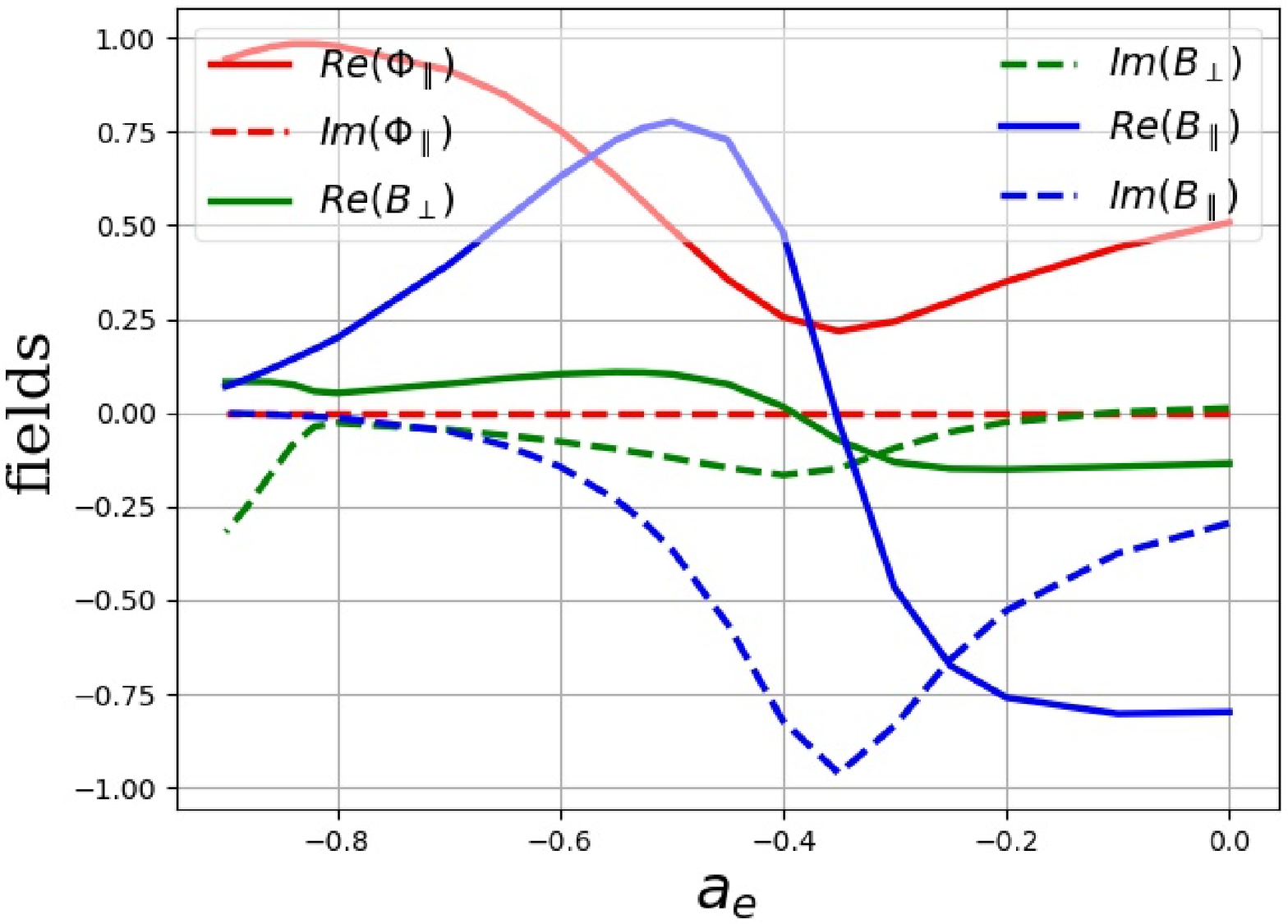}
\end{minipage}
}
	\caption{(Color online) Plots of wave polarization vs $a_{e}$. The other parameters are the same as in Fig. (\ref{eps:mm_ae_eig}).}
\label{eps:mm_pol_ae}
\end{figure*}

%\begin{linenumbers}
We then consider the additional effects of anisotropic electrons, i.e., finite $a_{e}$.
Figure (\ref{eps:mm_ae_eig}) displays the normal mode frequencies as $a_{e}$ varies.
Note that, with $a_{i}=-0.8$, the case considered here is ion mirror mode unstable at $a_{e}=0$.
It is found that, with decreasing $a_{e}$, there is a transition to the electron mirror mode, marked by a linear growth rate much higher than that associated with the ion mirror mode.
The ISW branch is further stabilized by $a_{e}$.
More interestingly, the higher eigenstates of the  MM branch, e.g., $M_{3}$,  move towards the marginal stability as $a_{e}$ decreases.
KAW can then be coupled to the mirror branch.
Specifically, for a strong drive case with $a_{e}+1\ll a_{i}+1\ll 1$,   Eq. (\ref{eq:emmdis}) indicates that the ordering estimate for the complex frequency of (electron) mirror mode is $|\xi_{e}Z_{e}+1|\sim \mathcal{O}(a_{e}+1)$.  Therefore, by taking the first order FILR correction into account, the dispersion relation, Eq. (\ref{eq:char}), can be reduced to the following form
\begin{eqnarray}
\label{eq:shortchar}
	(\Omega^{2}V_{1}+V_{2})[1+\beta_{i\perp}+\beta_{e\perp}-\frac{\beta_{e\perp}(1+\xi_{e}Z_{e})}{2(1+a_{e})}]=\mathcal{O}(b_{i}),
\end{eqnarray}
Equation (\ref{eq:shortchar}), thus, demonstrates  the coupling between KAW and electron MM due to the  FILR effect.
The wave polarization of $M_{0}$ mirror mode is given in Fig.(\ref{eps:mm_pol_ae}a). We see that the direction of parallel electric potential changes as $a_{e}$ decreases, and the wave polarization is dominated by $\Phi_{\parallel}$ in the electron mirror mode regime, consistent with the theoretical result, i.e., Eq. (\ref{eq:emmpol}).
Due to the KAW-MM coupling, another weakly damped mirror mode $M_{3}$ appears. Unlike $M_{0}$, this mode has a more significant perpendicular magnetic field perturbation in the small electron anisotropy regime, as shown in Fig. (\ref{eps:mm_pol_ae}d). 
The wave polarization of ion-sound wave is nearly independent of $a_{e}$.
As a result of coupling with the high-order normal modes of the mirror branch, the variation in $a_{e}$ is observed to strongly influence the wave polarization of KAW, as shown in Fig.(\ref{eps:mm_pol_ae}c). 
While the perpendicular field $B_{\perp}$ dominates when the magnitude of $|a_{e}|$ is small, the scalar parallel potential $\Phi_{\parallel}$  becomes dominant in the large $|a_{e}|$ regime.
However, it is worthwhile noting that here the $\Phi_{\parallel}$ is due to the coupling to MM branch, not the ISW branch. 
%\end{linenumbers}

\subsection{Firehose instability regime}
\begin{figure}[!htp]
\vspace{-0.3cm}
\setlength{\belowcaptionskip}{-0.1cm}
\centering
\includegraphics[scale=0.40]{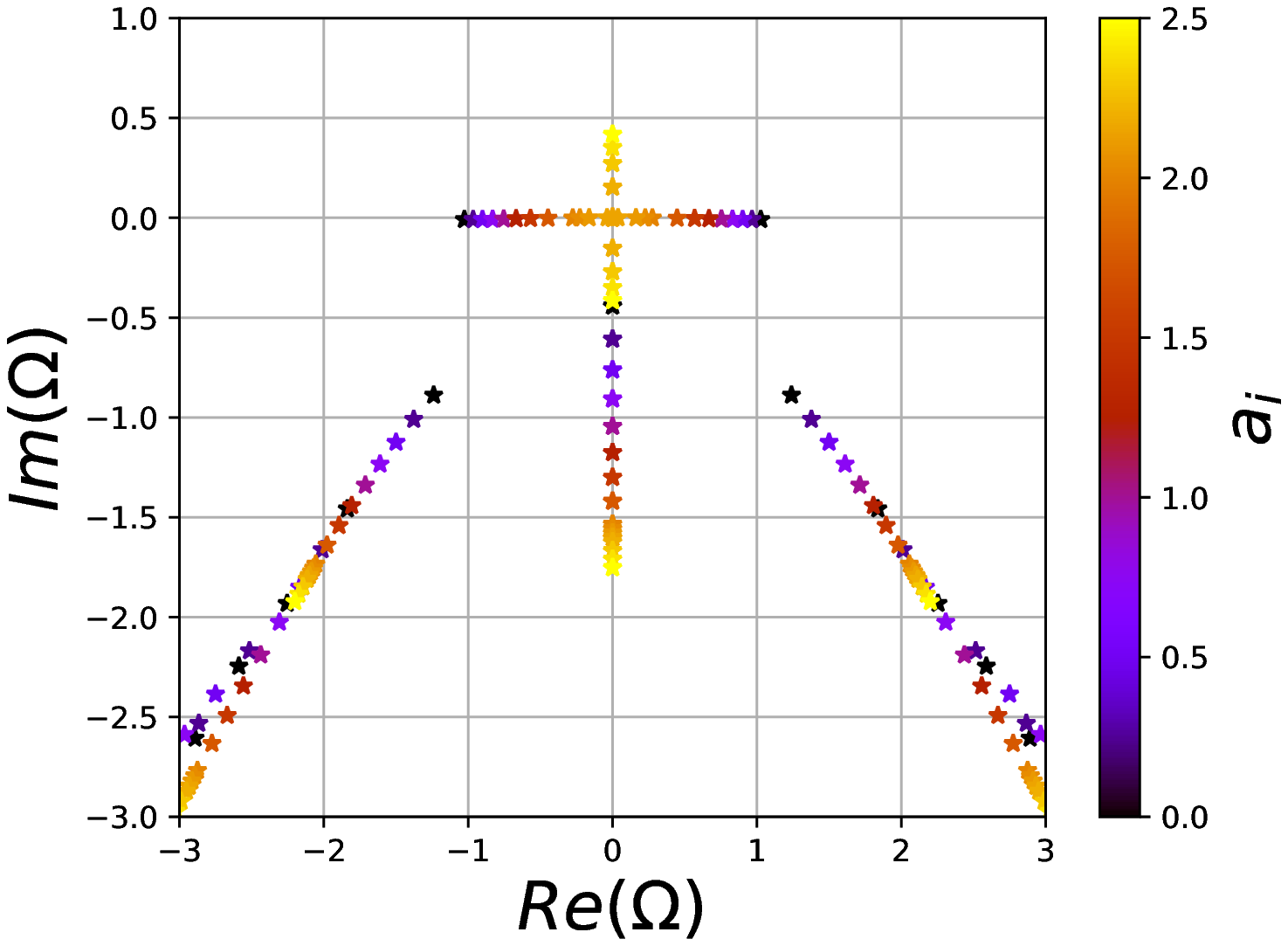}
	\caption{(Color online) Normalized complex frequency $\Omega$ vs $a_{i}$  in the region $-3\le Im(\Omega)\le 1$ and $-3\le Re(\Omega)\le 3$, for $b_{i}=0.1$, $\beta_{i\perp}=1$, $\tau=1$, $a_{e}=0$ and $m_{i}/m_{e}=1836$.}
\label{eps:fh_ai_eig}
\end{figure}
\begin{figure*}[!htp]
\centering
	\subfloat[\small{MM}]{
\label{eps:fh_ai_pol_mm}
\begin{minipage}[t]{0.32\textwidth}
\centering
\includegraphics[scale=0.30]{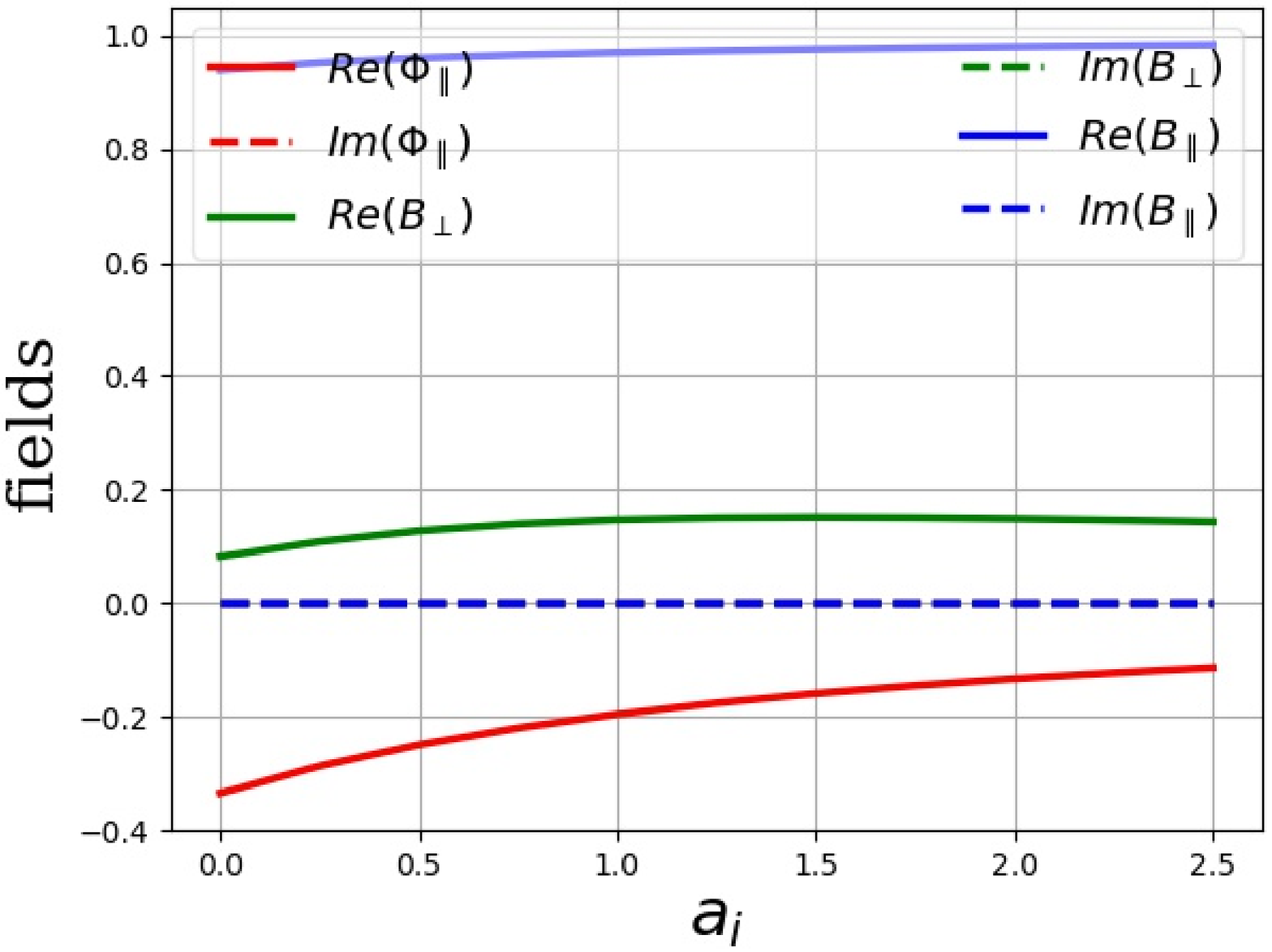}
\end{minipage}
}
	\subfloat[\small{ISW}]{
\label{eps:fh_ai_pol_is}
\begin{minipage}[t]{0.32\textwidth}
\centering
\includegraphics[scale=0.30]{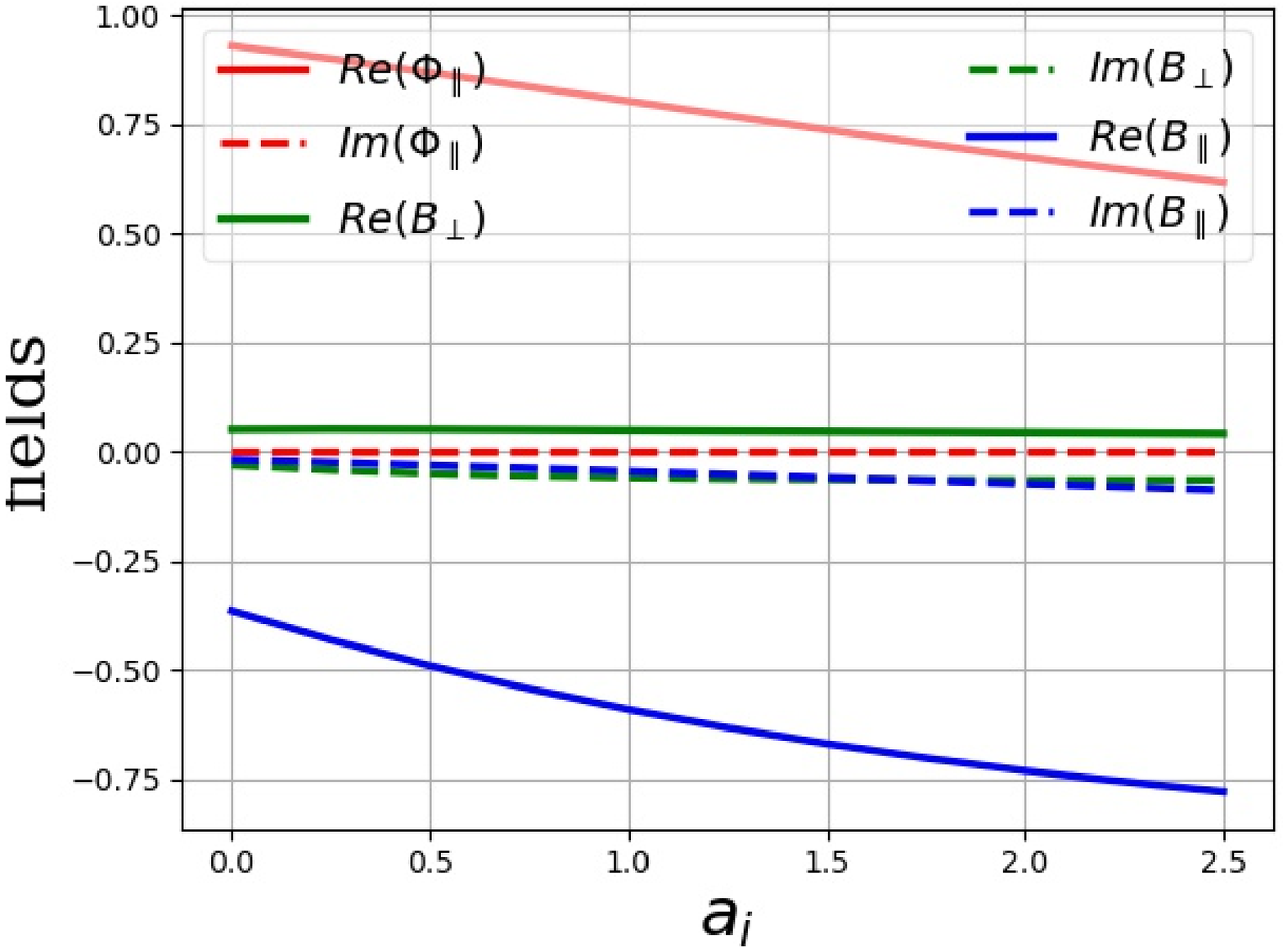}
\end{minipage}
}
	\subfloat[\small{KAW}]{
\label{eps:fh_ai_pol_al}
\begin{minipage}[t]{0.32\textwidth}
\centering
\includegraphics[scale=0.30]{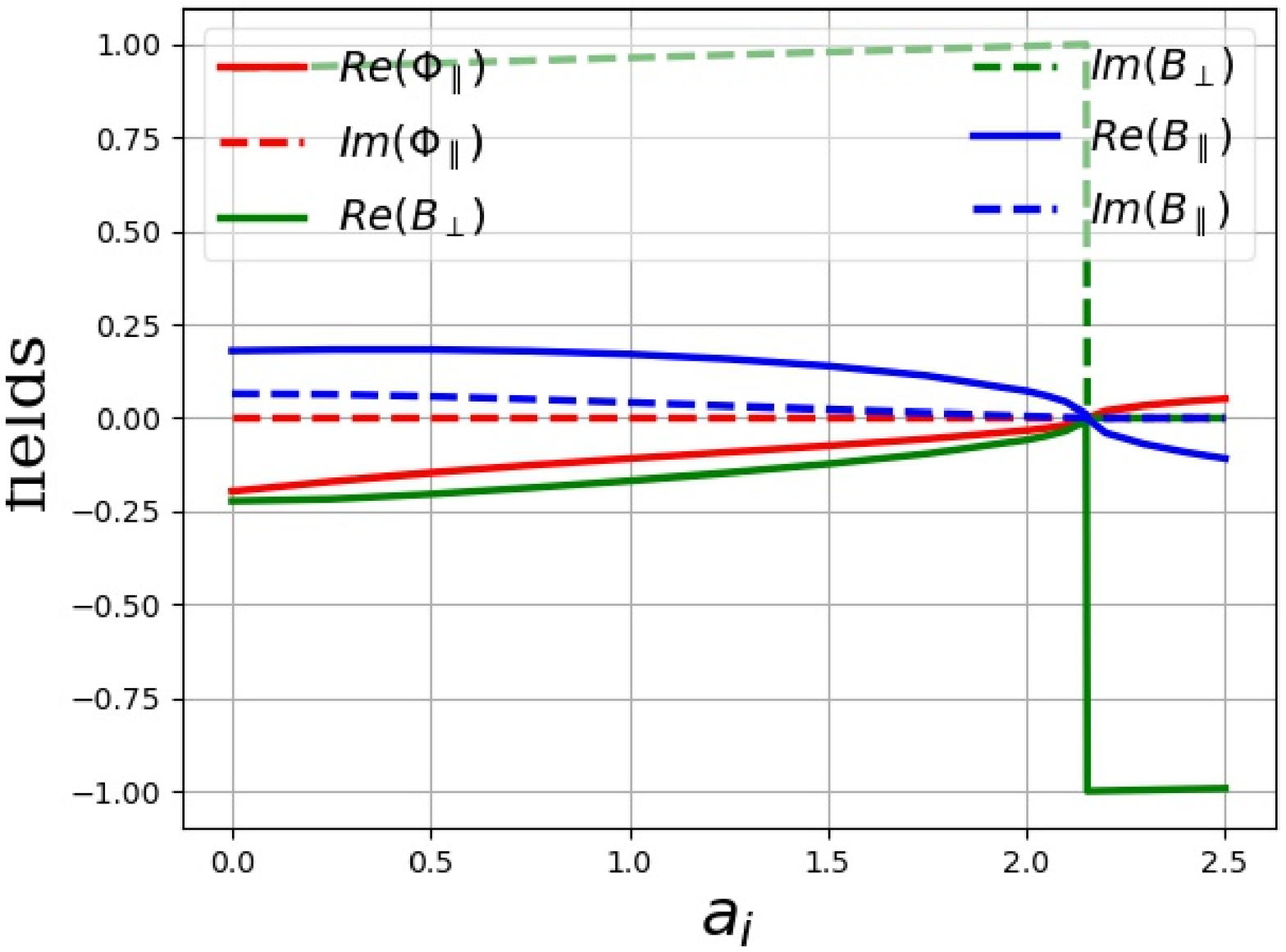}
\end{minipage}
}
	\caption{(Color online) Plots of wave polarization vs $a_{i}$. The other parameters are the same as in Fig. (\ref{eps:fh_ai_eig}).}
\label{eps:fh_pol_ai}
\end{figure*}
\begin{figure}[!htp]
\vspace{-0.3cm}
\setlength{\belowcaptionskip}{-0.1cm}
\centering
\includegraphics[scale=0.40]{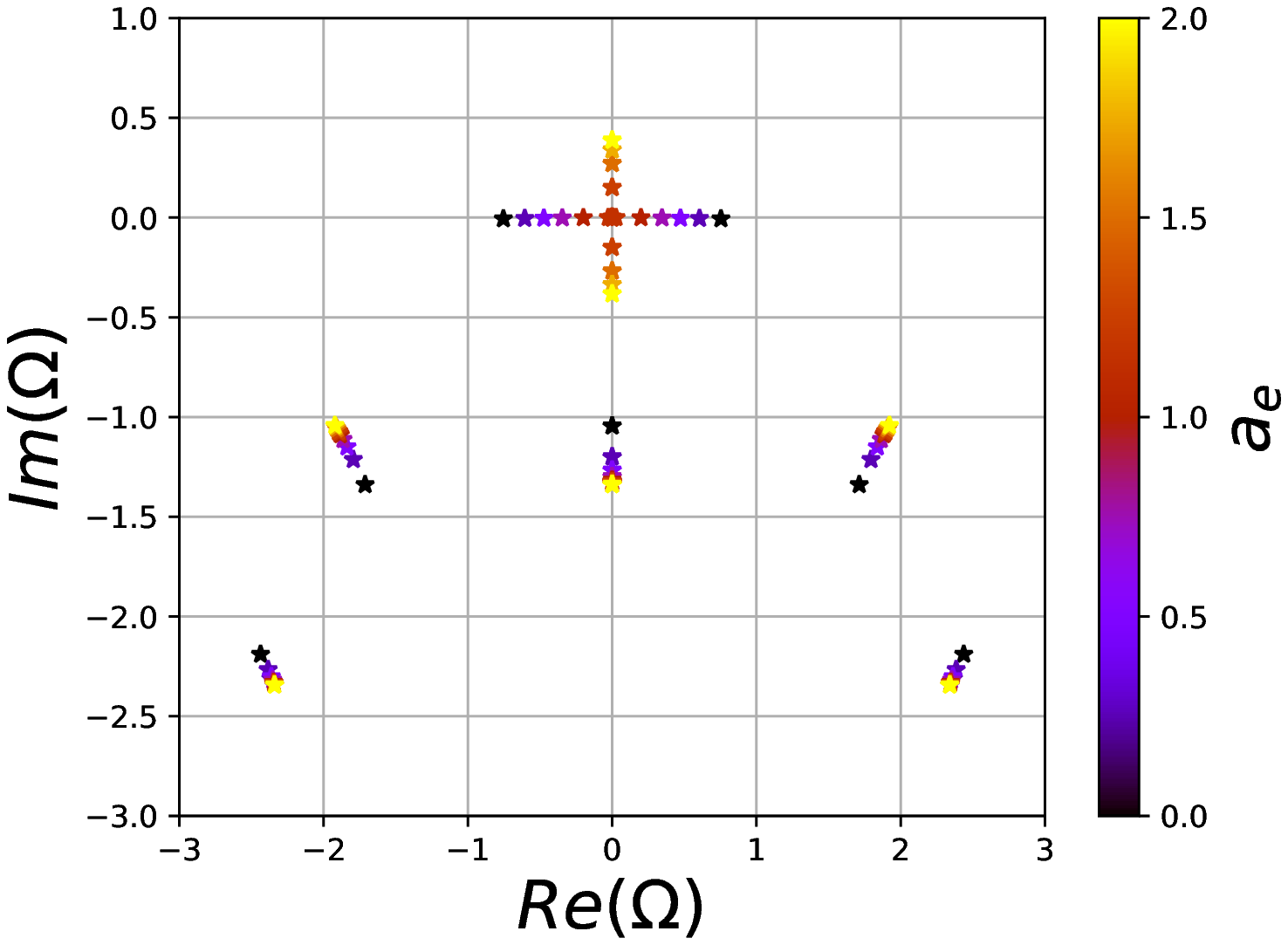}
	\caption{(Color online) Normalized complex frequency $\Omega$ vs $a_{e}$  in the region $-3\le Im(\Omega)\le 1$ and $-3\le Re(\Omega)\le 3$, for $b_{i}=0.1$, $\beta_{i\perp}=1$, $\tau=1$, $a_{i}=1$ and $m_{i}/m_{e}=1836$.}
\label{eps:fh_ae_eig}
\end{figure}
\begin{figure*}[!htp]
\centering
	\subfloat[\small{MM}]{
\label{eps:fh_ae_pol_mm}
\begin{minipage}[t]{0.32\textwidth}
\centering
\includegraphics[scale=0.30]{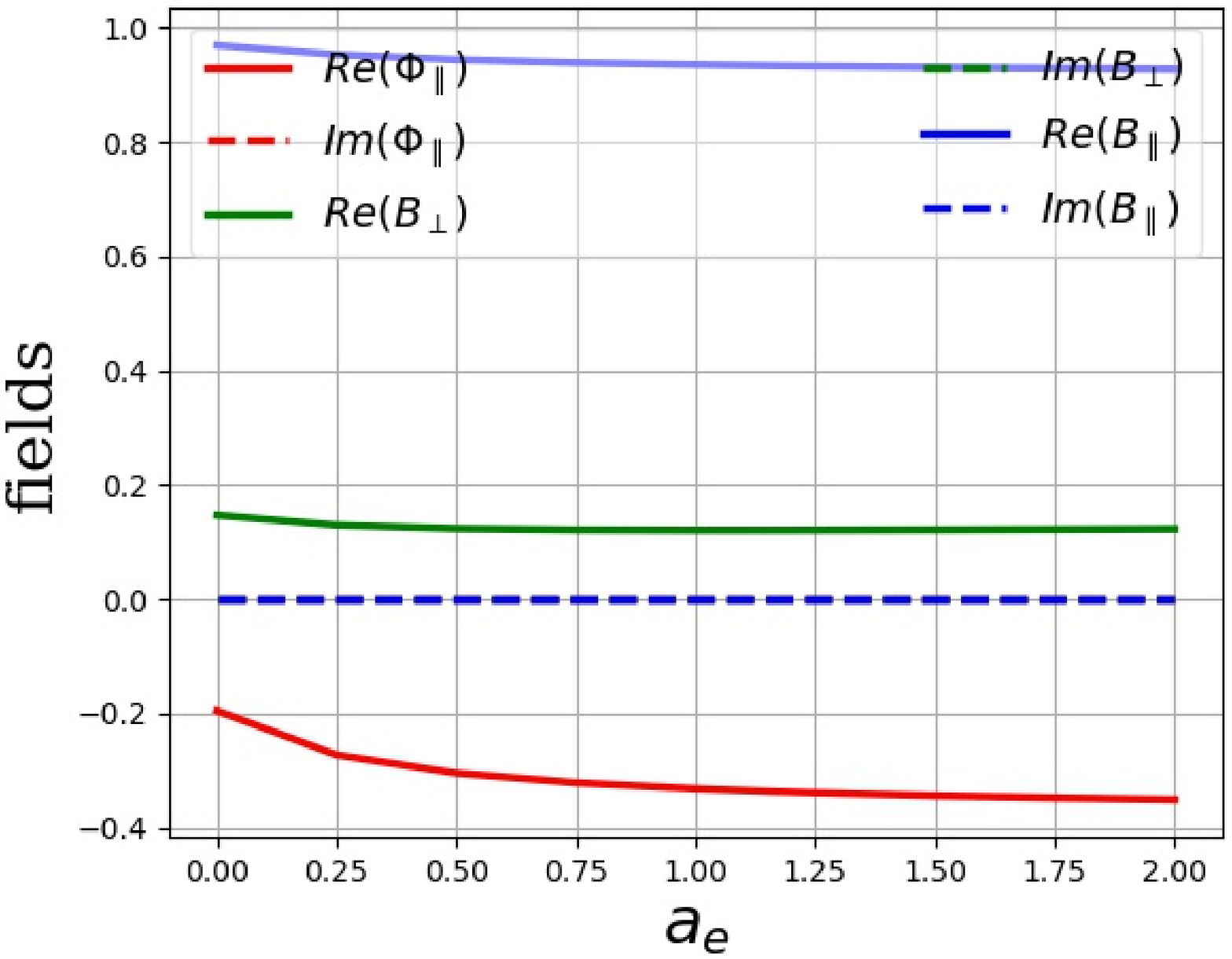}
\end{minipage}
}
	\subfloat[\small{ISW}]{
\label{eps:fh_ae_pol_is}
\begin{minipage}[t]{0.32\textwidth}
\centering
\includegraphics[scale=0.30]{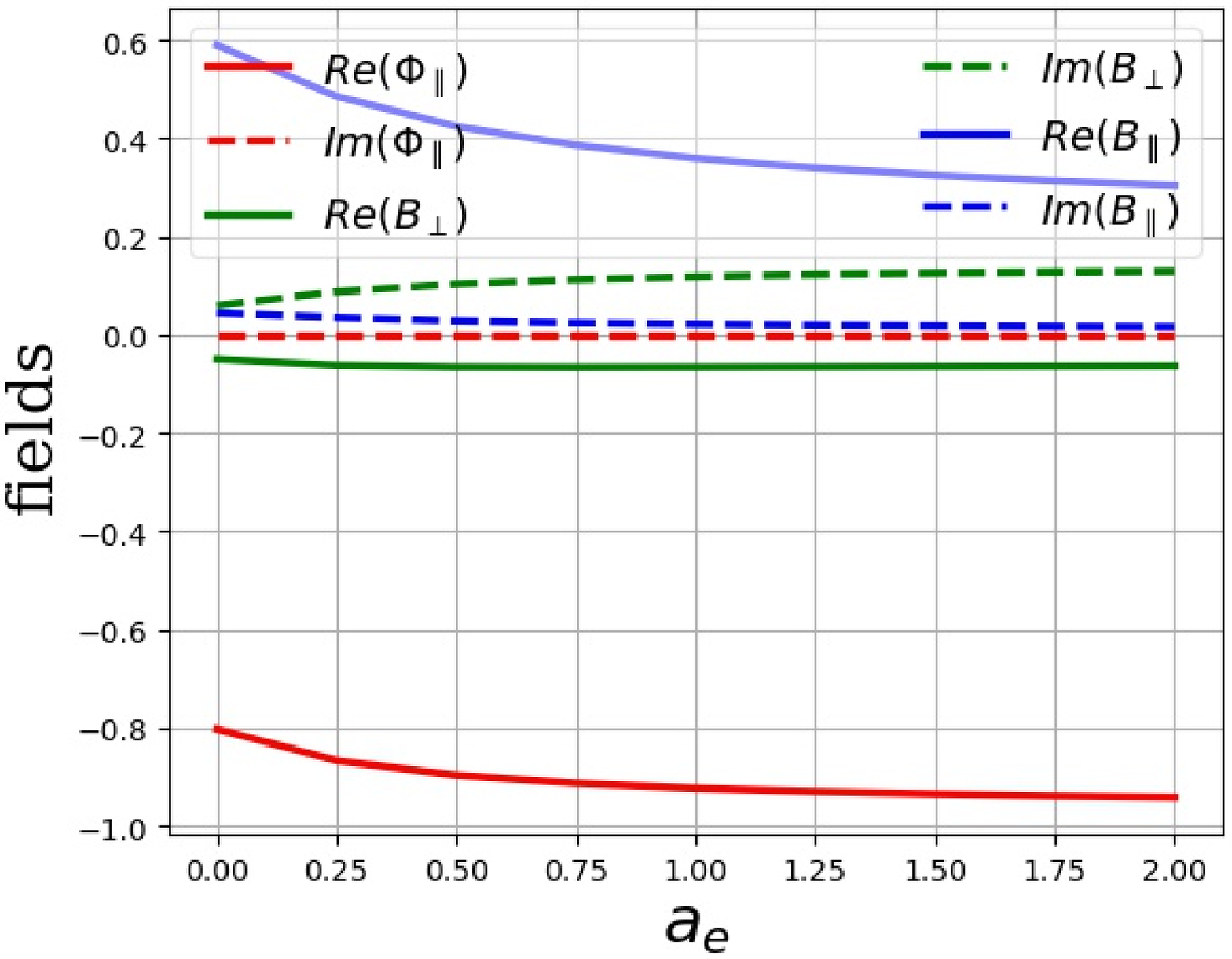}
\end{minipage}
}
	\subfloat[\small{KAW}]{
\label{eps:fh_ae_pol_al}
\begin{minipage}[t]{0.32\textwidth}
\centering
\includegraphics[scale=0.30]{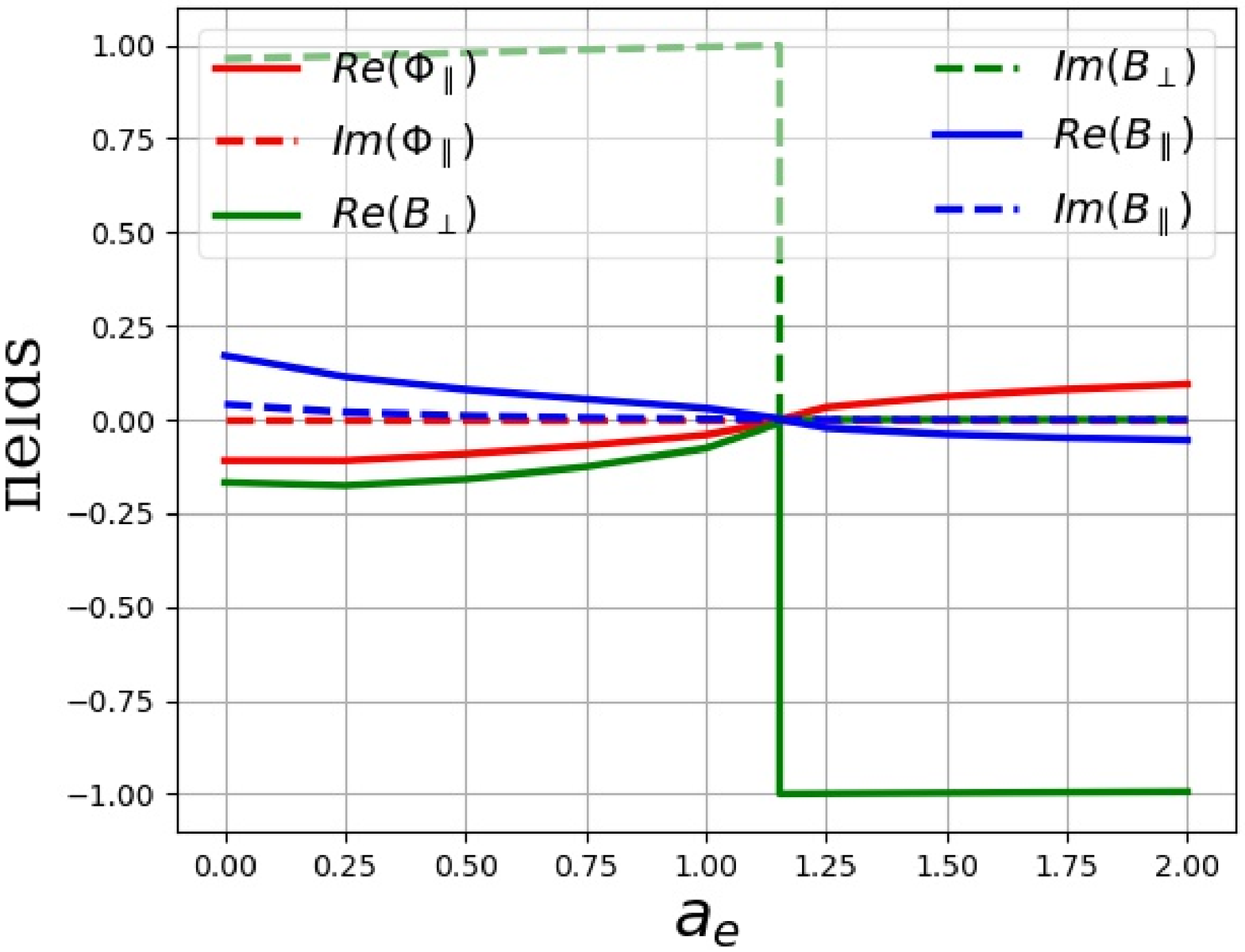}
\end{minipage}
}
	\caption{(Color online) Plots of wave polarization vs $a_{e}$. The other parameters are the same as in Fig. (\ref{eps:fh_ae_eig}).}
\label{eps:fh_pol_ae}
\end{figure*}
%\begin{linenumbers}
In addition to the mirror instability, another well-known instability in anisotropic plasma  is the firehose instability which excites SAWs/KAWs. As shown by  Eq. (\ref{eq:firehose}), it can arise in a high-$\beta$ anisotropic plasma with $a_{s}>0$ ($T_{s,\parallel}>T_{s\perp}$).
Since, from Eq. (\ref{eq:firehose}), the anisotropy drive, $a_{s}$, is additive, it follows that
 the ion driven firehose instability could be excited with $a_{i}>0$ and $a_{e}=0$.
The normal mode frequencies are shown for a range of $a_{i}$ values in Fig. (\ref{eps:fh_ai_eig}). For $a_{i}>0$, a significant stabilization effect on mirror and ion-sound branches is observed.  
As the instability drive, $a_{i}$, for KAW increases, the two weakly damped normal modes approaches the marginal stability point ($\Omega=0$) to form a solution of multiplicity $2$, it then splits into two simple solutions, a stable and an unstable, when $a_{i}$ exceeds the threshold. 
The normal mode frequencies of KAW ultimately remain in the imaginary axis. 
This feature is in agreement with the general theoretical analysis given in Sec. (\ref{sec:oct:15:08:18}).
The corresponding wave polarizations for varying $a_{i}$ are illustrated in Fig. (\ref{eps:fh_pol_ai}).
We note that, as $a_{i}$ increases, $|\Phi_{\parallel}/B_{\parallel}|$ of both MM and ISW branches decrease.
The KAW branch, meanwhile, is still dominated by the perpendicular magnetic fluctuation throughout the $a_{i}$ variation. However, $ B_{\perp}$ is  discontinuous at the $a_{i}$ threshold, where marginal stability occurs at $\Omega=0$ and undergoes $\pi/2$ phase shift.

The normal mode frequencies varying with $a_{e}$ are shown in Fig. (\ref{eps:fh_ae_eig}).
The stability property of KAW with $a_{e}$ drive, as expected, is similar to that of $a_{i}$ drive.
$a_{e}$, meanwhile, has relatively weaker effects on the damping rates of ISW and MM.
Figure (\ref{eps:fh_pol_ae}) shows wave polarizations, which are similar to those with $a_{i}$ variations.
In contrast to the mirror instability regime, here the wave polarizations of KAW and MM show weak dependence on $a_{e}$.

\section{Conclusions and Discussions}
\label{sec:oct:01:16:05}
In this paper, we have employed the gyrokinetic theory and studied the linear stability and polarization properties of low-frequency electromagnetic fluctuations in finite-$\beta$ anisotropic uniform plasmas. 
Consistent with the gyrokinetic ordering, the current model is valid for wavelengths from larger than to comparable with the microscopic thermal electron Larmor radius,  and, thus,  provides a self-consistent kinetic description of the ion-sound wave, shear/kinetic Alfv\'{e}n wave, and mirror mode branches.

The behaviour of eigenvalues; namely the complex normal mode frequencies of an analytically derived linear dispersion relation, is discussed in detail.
Specifically, for a bi-Maxwellian plasma,  we obtain a general form of kinetic stability criteria analytically for the firehose and mirror instabilities, including effects of finite Larmor radius and wave-particle interactions. 
It can thus be used to quantitatively identify the excitation mechanisms  and set anisotropy constraints for fluctuations from both satellite observations and numerical simulation results. 
Moreover, in order to clarify the distinctive features of  different branches and gain necessary insights into the underlying linear physics, the governing equations are further examined in both the long-wavelength and low-$\beta$ limits..
In the low-$\beta$ limit, it recovers the well-known result that the kinetic Alfv\'{e}n wave arises from the coupling to ion-sound branch due to the FILR effect.
In the long-wavelength limit, the mirror mode branch is characterized by a strong coupling to the ion-sound wave branch, and, thereby, both the normal mode frequency and wave polarization are modified significantly.

Further extensive studies have been carried out numerically by using a novel eigenvalue-solver. We present here a broad spectrum of low-frequency normal modes, and perform systematic investigations  on the corresponding linear wave properties  over a set of parameters.
In particular, it is demonstrated that while the mode frequencies of the ion-sound wave branch can be significantly influenced by the electron to ion temperature ratio, $\tau$, the wave polarizations are insensitive to it.
For the kinetic Alfv\'{e}n wave and mirror mode branches, however, $\tau$ mainly changes their wave polarizations.
Finite ion Larmor radius, $b_{i}$, is generally stabilizing for the low-frequency fluctuations, and provides intrinsic couplings among the three wave branches.
Positive (negative) anisotropy, $a_{s}$, excites the firehose (mirror) instability. In the mirror instability regime, the increasing of electron anisotropy can lead to a significant coupling between the kinetic Alfv\'{e}n wave and high-order mirror modes, and  affect the wave frequency and wave polarization of KAW. In particular, consistent with the analytical theory, the wave polarization of the electron mirror mode is dominated by $\Phi_{\parallel}$.
In the firehose instability regime, the kinetic Alfv\'{e}n wave becomes unstable via a reactive process, and the dominant wave characteristic remains to be the perpendicular magnetic field fluctuation.

Finally,  we remark that the present work is limited to uniform plasmas in order to illustrate the detailed wave and stability properties and their dependences on the physical parameters. It is obvious and desirable to extend the present analysis to nonuniform plasmas including, e.g., effects of diamagnetic drifts and instability drives due to the density and/or temperature gradients. 
These studies will be reported in future publications.
%it should be noted that the present model can be further extended to include realistic equilibrium geometry and spatial nonuniformities. 
%However, the formulation of such a general problem is beyond the intended scope of this work and will be discussed in future publications.

%\section{Acknowledgments}
%\label{sec:may:01:16:05}
\acknowledgments

One of the authors (H. T. Chen) would like to thank F. Zonca, E. Viezzer, M. Gracia-Munoz, M. Zhou and Z. Qiu for useful conversations.
This work was supported by National Natural Science Foundation of China under Grant Nos. 11905097 and 11235009.
The support from the European Research Council (ERC) under the European Unions Horizon 2020 research and innovation programme (grant agreement No. 805162) is gratefully acknowledged.
%\end{linenumbers}

\section*{Data Availability}
The data that support the findings of this study are available from the corresponding author upon reasonable request.

\bibliography{main}

\end{document}